\newcommand{\diracslash}[1]{#1\llap{/\kern2pt}}

\newcommand{\be}{\begin{equation}}
\newcommand{\ee}{\end{equation}}
\newcommand{\bea}{\begin{eqnarray}}
\newcommand{\eea}{\end{eqnarray}}
\newcommand{\ba}[1]{\begin{array}{#1}}
\newcommand{\ea}{\end{array}}

\newcommand{\bt}{\begin{tabular}}
\newcommand{\et}{\end{tabular}}

\newcommand{\beas}{\begin{eqnarray*}}
\newcommand{\eeas}{\end{eqnarray*}}

\documentclass[preprint,prd,aps,floats,nofootinbib,showpacs,floatfix]{revtex4}
\usepackage{graphicx}
\begin{document}

\title{D mesons and charmonium states in asymmetric nuclear matter
at finite temperatures}
\author{Arvind Kumar}
\email{iitd.arvind@gmail.com}
\affiliation{Department of Physics, Indian Institute of Technology, Delhi,
Hauz Khas, New Delhi -- 110 016, India}

\author{Amruta Mishra}
\email{amruta@physics.iitd.ac.in,mishra@th.physik.uni-frankfurt.de}
\affiliation{Department of Physics, Indian Institute of Technology, Delhi,
Hauz Khas, New Delhi -- 110 016, India}

\begin{abstract}
We investigate the in-medium masses of $D$ and $\bar{D}$ mesons in 
the isospin-asymmetric nuclear matter at finite temperatures 
arising due to the interactions with the nucleons, the scalar 
isoscalar meson $\sigma$, and the scalar iso-vector meson $\delta$
within a SU(4) model. However, since the chiral symmetry is explicitly 
broken for the SU(4) case due to the large charm quark mass, we use 
the SU(4) symmetry here only to obtain the interactions of the $D$ 
and $\bar D$ mesons with the light hadron sector, but use the observed 
values of the heavy hadron masses and empirical values of the decay 
constants. The in-medium masses of $J/\psi$ and the excited charmonium 
states ($\psi(3686)$ and $\psi(3770)$) are also calculated in the hot 
isospin asymmetric nuclear matter in the present investigation. These 
mass modifications arise due to the interaction of the charmonium states 
with the gluon condensates of QCD, simulated by a scalar dilaton field 
introduced to incorporate the broken scale invariance of QCD within the
effective chiral model. The change in the mass of $J/\psi$ in 
the nuclear matter with the density is seen to be rather small, 
as has been shown in the literature by using various approaches, 
whereas, the masses of the excited states of charmonium ($\psi(3686)$ 
and $\psi(3770)$) are seen to have considerable drop at high densities.
The present study of the in-medium masses of $D$ ($\bar{D}$) mesons 
as well as of the charmonium states will be of relevance for the 
observables from the compressed baryonic matter, like the production 
and collective flow of the $D$ ($\bar D$) mesons, resulting from the 
asymmetric heavy ion collision experiments planned at the future 
facility of the FAIR, GSI. The mass modifications of $D$ and $\bar{D}$ 
mesons as well as of the charmonium states in hot nuclear medium can  
modify the decay of the charmonium states ($\Psi^{'}, \chi_{c}, J/\Psi$) 
to $D\bar{D}$ pairs in the hot dense hadronic matter. The small attractive 
potentials observed for the $\bar{D}$ mesons may lead to formation 
of the $\bar{D}$ mesic nuclei. 

\end{abstract}
\maketitle

\def\bfm#1{\mbox{\boldmath $#1$}}

\section{Introduction}
The study of in-medium properties of hadrons is important for 
the understanding of strong interaction physics. The study of 
in-medium hadron properties has relevance in heavy-ion collision
experiments as well as in nuclear astrophysics. There have been also
extensive experimental efforts for the study of in-medium hadron 
properties by nuclear collision experiments. In these heavy-ion 
collision experiments, hot and dense matter is produced. By studying 
the experimental observables one can infer about how the hadron 
properties are modified in the medium. For example, the observed 
enhanced dilepton spectra \cite{ceres,helios,dls} could be a 
signature of medium modifications of the vector mesons 
\cite{Brat1,CB99,vecmass,dilepton,liko}.
Similarly the properties of the kaons and antikaons have been studied 
experimentally by KaoS collaboration and the production of kaons and 
antikaons in the heavy-ion collisions and their collective flow are 
directly related to the medium modifications of their spectral 
functions \cite{CB99,cmko,lix,Li2001,K5,K6,K4,kaosnew}. The study 
of $D$ and $\bar{D}$ mesons properties will be of direct relevance
for the upcoming experiment at FAIR, GSI, where one expects to produce
matter at high densities and moderate temperatures \cite{gsi}.
At such high densities, the properties of the $D$ and $\bar{D}$ mesons 
produced in these experiments are expected to be modified which should
reflect in experimental observables like their production and propagation
in the hot and dense medium. The reason for an expected appreciable
modifications of the $D$ and $\bar D$ mesons is
that $D$ and $\bar{D}$ mesons contain a light quark (u,d) 
or light antiquark. This light quark or antiquark interacts 
with the nuclear medium and leads to the modifications of $D$ 
and $\bar{D}$ properties. 
The experimental signature for this can be their production ratio 
and also in-medium $J/\psi$ suppression \cite{NA501,NA50e,NA502}. 
In heavy-ion collision experiments of much higher collision energies,
for example in RHIC or LHC, it is suggested that the $J/\psi$ 
suppression is because of the formation of quark-gluon plasma 
(QGP) \cite{blaiz,satz}. However, in Ref. \cite{zhang,brat5,elena} 
it is observed that the effect of hadron absorption of $J/\psi$ 
is not negligible. 
In Ref.\cite{vog}, it was reported that the charmonium suppression 
observed in Pb + Pb collisions of NA50 experiment cannot be simply explained
by nucleon absorption, but needs some additional density dependent suppression 
mechanism. It was suggested in these studies that the comover scattering
\cite{vog,capella,cassing} can explain the additional suppression of 
charmonium. An important difference between $J/\psi$ suppression pattern 
in comovers interaction model and in a deconfining scenario is that, in the 
former case, the anomalous suppression sets in smoothly from peripheral 
to central collisions rather than in a sudden way when the deconfining 
threshold is reached \cite{capella}.
The $J/\psi$ suppression in nuclear collisions at SPS energies has been 
studied in covariant transport approach HSD in Ref.\cite{cassing}. The 
calculations show that the absorption of $J/\psi$'s by both nucleons and 
produced mesons can explain reasonably not only the total $J/\psi$ 
cross--section but also the transverse energy dependence of $J/\psi$ 
suppression measured in both proton-nucleus and nucleus collisions. 
In Ref.\cite{wang}, the cross section of $J/\psi$ dissociation
by gluons is used to calculate the $J/\psi$ suppression in an equilibrating
parton gas produced in high energy nuclear collisions. The large average
momentum in the hot gluon gas enables gluons to break up the $J/\psi$, 
while hadron matter at reasonable temperature does not provide sufficiently 
hard gluons. 

Due to the reduction in the masses of $D$ and 
$\bar{D}$ mesons in the medium it is a possibility that excited 
charmonium states can decay to $D\bar{D}$ pairs \cite{brat6} 
instead of decaying to lowest charmonium state $J/\psi$. 
Actually higher charmonium states are considered as major 
source of $J/\psi$ \cite{pAdata}. Even at certain higher densities,
it can become a possibility that the $J/\psi$ itself will decay 
to $D\bar{D}$ pairs. So this can be an explanation of the observed 
$J/\psi$ suppression by NA50 collaboration at $158$ GeV/nucleon 
in the Pb-Pb collisions \cite{blaiz}. The excited states of 
charmonium also undergo mass drop in the nuclear medium \cite{leeko}.
The modifications of the in-medium masses of $D$ mesons is large
then the $J/\psi$ mass modification \cite{haya1,friman}.
This is because the charmonium states are made up of a heavy charm quark and
a charm antiquark. Within QCD sum rules, it is suggested that these 
heavy charmonium states interact with the nuclear medium through 
the gluon condensates. This is contrary to the interaction of the 
light vector mesons ($\rho, \omega, \phi$), which interact 
with the nuclear medium through the quark condensates. This is 
because all the heavy quark condensates can be related to the 
gluon condenstaes via heavy-quark expansion \cite{kimlee}. 
Also in the nuclear medium there are no valence charm quark 
to leading order in density and any interaction with the medium 
is gluonic. The QCD sum rule approach \cite{klingl} and leading order 
perturbative calculations \cite{pes1} to study the medium modifications 
of charmonium, show that the mass of $J/\psi$ is reduced slightly 
in the nuclear medium. In \cite{leeko}, the mass modification of charmonium
has been studied using leading order QCD formula and the linear 
density approximation for the gluon condensate in the nuclear medium.
This shows a small drop for the $J/\psi$ mass at the nuclear matter 
density, but there is seen to be significant shift in the masses of 
the excited states of charmonium ($\psi(3686)$ and $\psi(3770)$).

The in-medium modifications of $D$ and $\bar{D}$ mesons have been studied 
using various approaches. For example, in the QCD sum rule approach,
it is suggested 
that the light quark or antiquark of $D (\bar D)$ mesons interacts with
the light quark condensate leading to the medium modification of the 
$D(\bar D)$ meson masses \cite{arata,qcdsum08}. The quark meson coupling 
(QMC) model has also been used to study the D-meson properties \cite{qmc}. 
In the QMC model, the light quarks (u,d) and light antiquarks 
($\bar u$,$\bar d$)
confined in the nucleons and mesons interact via exchange of a 
scalar-isoscalar $\sigma$ meson as well a vector $\omega$ meson. 
The nucleon has a large reduction of mass in the dense medium arising 
due to the interaction of the light quarks (u,d) with the $\sigma$ field. 
The drop in the mass of $D$ mesons 
observed in the QMC model turns out to be similar to those calculated 
within the QCD sum rule approach. 

In the present investigation, we study the properties of the $D$ and $\bar{D}$ 
mesons in the isospin-asymmetric hot nuclear matter. These modifications
arise due to their interactions with the nucleons, the non-strange scalar 
isoscalar meson $\sigma$ and the scalar isovector meson $\delta$. 
We also study the medium modification of the masses of $J/\psi$ and 
excited charmonium states $\psi(3686)$ and $\psi(3770)$ in the nuclear 
medium due to the interaction with the gluon condensates using the 
leading order QCD formula. The gluon condensate in the nuclear medium 
is calculated from the medium modification of a scalar dilaton field 
introduced within a chiral SU(3) model \cite{paper3} through a scale 
symmetry breaking term in the Lagrangian density leading to the QCD trace 
anomaly. In the chiral SU(3) model, the gluon condensate is related 
to the fourth power of the dilaton field $\chi$ and the changes in 
the dilaton field with the density are seen to be small. We study the 
isospin dependence of the in-medium masses of charmonium obtained 
from the dilaton field, $\chi$ calculated for the asymmetric nuclear
matter at finite temperatures. The medium modifications of the light
hadrons (nucleons and scalar mesons) are described by using a chiral $SU(3)$ 
model \cite{paper3}. The model has been used to study finite nuclei, 
the nuclear matter properties, the in-medium properties of the vector mesons 
\cite{hartree,kristof1} as well as to investigate the optical potentials 
of kaons and antikaons in nuclear matter \cite{kmeson,isoamss} and in 
hyperonic matter in \cite{isoamss2}. For the study of the properties $D$ 
mesons in isospin-asymmetric medium at finite temperatures, the chiral 
SU(3) model is generalized to $SU(4)$ flavor symmetry to obtain the 
interactions of $D$ and $\bar{D}$ mesons with the light hadrons. 
Since the chiral symmetry is explicitly broken for the SU(4) case
due to the large charm quark mass, we use the SU(4) symmetry here
only to obtain the interactions of the $D$ and $\bar D$ mesons with
the light hadron sector, but use the observed values of the heavy
hadron masses and empirical values of the decay constants. This 
has been in line with the philosophy followed in Ref. \cite{liukolin}
where charmonium absorption in nuclear matter was studied using 
the SU(4) model to obtain the relevant interactions. However,
the values of the heavy hadron masses and the coupling constants 
in Ref. \cite{liukolin}, were taken as the empirical values 
or as calculated from other theoretical models. The coupling
constants were derived by using the relations from SU(4) symmetry, 
if neither the empirical values nor values calculated from other 
theoretical models were available \cite{liukolin}.
The $D$ meson properties in symmetric hot nuclear matter using SU(4)
model have been studied in ref. \cite{amdmeson} and for the 
case of asymmetric nuclear matter at zero temperature in \cite{amarind}. 
In a coupled channel approach  for the study of D mesons,
using a separable potential, it was shown that the resonance $\Lambda_c (2593)$
is generated dynamically in the I=0 channel \cite{ltolos} analogous to
$\Lambda (1405)$ in the coupled channel approach for the $\bar K N$ interaction
\cite{kbarn}. The approach has been generalized to study the spectral density 
of the D-mesons at finite temperatures and densities \cite{ljhs},
taking into account the modifications of the nucleons in the medium. 
The results of this investigation seem to indicate a dominant increase 
in the width of the D-meson whereas there is only a very small change
in the D-meson mass in the medium \cite{ljhs}. However, these calculations
\cite{ltolos,ljhs}, assume the interaction to be SU(3) symmetric in u,d,c 
quarks and ignore channels with charmed hadrons with strangeness.
A coupled channel approach for the study of D-mesons has been developed based
on SU(4) symmetry \cite{HL} to construct the effective interaction between 
pseudoscalar mesons in a 16-plet with baryons in 20-plet representation
through exchange of vector mesons and with KSFR condition \cite{KSFR}.
This model \cite{HL} has been modified in aspects like regularization method
and has been used to study DN interactions in Ref. \cite{mizutani6}. 
This reproduces the resonance  $\Lambda_c (2593)$ in the I=0 channel and in
addition generates another resonance in the I=1 channel at around 2770 MeV.
These calculations have been generalized to finite temperatures 
\cite{mizutani8} 
accounting for the in-medium modifications of the nucleons in a Walecka type
$\sigma-\omega$ model, to study the $D$ and $\bar D$ properties \cite{MK} 
in the hot and dense hadronic matter. At the nuclear matter density and for
zero temperature, these resonances ($\Lambda_c (2593)$ and $\Sigma_c (2770)$)
are generated $45$ MeV and $40$ MeV below their free space positions. 
However at finite temperature, e.g., at $T = 100$ MeV resonance positions 
shift to $2579$ MeV and $2767$ MeV for $\Lambda_{c}$  ($I = 0$) and 
$\Sigma_{c}$ ($I = 1$) respectively. Thus at finite temperature resonances 
are seen to move closer to their free space values. This is  
because of the reduction of pauli blocking factor arising due to the 
fact that fermi surface is smeared out with temperature. For $\bar{D}$ 
mesons in coupled channel approach a small repulsive mass shift is obtained. 
This will rule out of any possibility of charmed mesic nuclei 
\cite{mizutani8} suggested in the QMC model \cite{qmc}. 
But as we shall see in our investigation, we obtain a small attractive mass 
shift for $\bar{D}$ mesons which can give rise to the possibility
of the formation of charmed mesic nuclei. The study of $D$ meson 
self-energy in the nuclear matter is also helpful in understanding
the properties of the charm and the hidden charm resonances in the 
nuclear matter \cite{tolosra}. In coupled channel 
approach the charmed resonance $D_{s0}(2317)$ mainly couples to $DK$ system, 
while the $D_{0}(2400)$ couples to $D\pi$ and $D_{s}\bar{K}$. The hidden 
charm resonance couples mostly to $D\bar{D}$. Therefore any modification 
of $D$ meson properties in the nuclear medium will affect the properties 
of these resonances. In Ref.\cite{haid1, haid2}, the $\bar{D}N$ interactions 
at low energies have been studied using a meson exchange model ($\omega$
and $\rho$) in close analogy with the meson exchange model
for the KN interactions \cite{julich}, supplemented with a 
short-distance contribution from one-gluon exchange. 
The scattering lengths for the I=0 and I=1 channels for the
$\bar D N$ interactions arising from the one gluon exchange 
are seen to be very close to the values in Ref. \cite{MK}.
Generalizing the SU(4) models with vector meson exchange potentials
to SU(8) spin-flavor symmetry, that treats the heavy 
pseudoscalar and vector mesons on equal footing-as required by 
heavy-quark symmetry, in Ref. \cite{cgar}, the charmed baryon 
resonances that are generated dynamically have been studied 
within a unitary meson-baryon coupled-channel model.
Some of the resonances in this model are identified with the
recently observed baryon resonances.
In the present investigation, the $D(\bar D)$ energies are modified 
due to a vectorial Weinberg-Tomozawa, scalar exchange terms 
($\sigma$, $\delta$) as well as range terms \cite{isoamss,isoamss2}.
The isospin asymmetric effects among $D^0$ and $D^+$ in the doublet,
D$\equiv (D^0,D^+)$ as well as between $\bar {D^0}$ and $D^-$ in the
doublet, $\bar D \equiv (\bar {D^0},D^-)$ arise due to the scalar-isovector
$\delta$ meson, due to asymmetric contributions in the Weinberg-Tomozawa
term, as well as in the range term \cite{isoamss}. 

We organize the paper as follows. In section II, we give a brief 
introduction to the effective chiral $SU(3)$ model used to study 
the isospin asymmetric nuclear matter at finite temperatures, and
its extension to the $SU(4)$ model to derive the interactions
of the charmed mesons with the light hadrons. In section III, we present 
the dispersion relations for the $D$ and $\bar{D}$ mesons to be solved 
to calculate their optical potentials in the hot and dense hadronic matter.  
In secton IV, we show how the in-medium masses of the charmonium states 
$J/\psi$, $\psi(3686)$ and $\psi(3770)$ in the present investigation,
arise from the medium modification of the scalar dilaton field, 
introduced within the chiral model to incorporate broken scale invariance 
leading to QCD trace anomaly. 
Section V contains the results and discussions and finally,
in section VI, we summarize the results of present investigation 
and discuss possible outlook.

\section{The hadronic chiral $SU(3) \times SU(3)$ model }

We use a chiral $SU(3)$ model for the study of the light hadrons in the 
present investigation \cite{paper3}. The model is based on nonlinear 
realization of chiral symmetry \cite{weinberg,coleman,bardeen} and 
broken scale invariance \cite{paper3,hartree,kristof1}. 
The effective hadronic chiral Lagrangian contains the following terms
\begin{equation}
{\cal L} = {\cal L}_{kin}+\sum_{W=X,Y,V,A,u} {\cal L}_{BW} + 
{\cal L}_{vec} + {\cal L}_{0} + {\cal L}_{SB}
\end{equation}
In Eq.(1), ${\cal L}_{kin}$ is the kinetic energy term, ${\cal L}_{BW}$ 
is the baryon-meson interaction term in which the baryons-spin-0 meson 
interaction term generates the baryon masses. ${\cal L}_{vec}$  describes 
the dynamical mass generation of the vector mesons via couplings to the 
scalar mesons and contain additionally quartic self-interactions of the 
vector fields. ${\cal L}_{0}$ contains the meson-meson interaction terms 
inducing the spontaneous breaking of chiral symmerty as well as a scale 
invariance breaking logarthimic potential. ${\cal L}_{SB}$ describes the 
explicit chiral symmetry breaking.

To study the hadron properties at finite temperature and densities
in the present investigation, we use the mean  field approximation,
where all the meson fields are treated as classical fields. 
In this approximation, only the scalar and the vector fields 
contribute to the baryon-meson interaction, ${\cal L}_{BW}$
since for all the other mesons, the expectation values are zero.
The interactions of the scalar mesons and vector mesons with the
baryons are given as
\begin{eqnarray}
{\cal  L} _{Bscal} +  {\cal L} _{Bvec} = - \sum_{i} \bar{\psi}_{i} 
\left[  m_{i}^{*} + g_{\omega i} \gamma_{0} \omega 
+ g_{i\rho} \gamma_{0} \rho + g_{\phi i} \gamma_{0} \phi 
\right] \psi_{i}. 
\label{lagscvec}
\end{eqnarray}
The interaction of the vector mesons, of the scalar fields and 
the interaction corresponding to the explicitly symmetry breaking
in the mean field approximation are given as
\begin{eqnarray}
 {\cal L} _{vec} & = & \frac{1}{2} \left( m_{\omega}^{2} \omega^{2} 
+ m_{\rho}^{2} \rho^{2} + m_{\phi}^{2} \phi^{2} \right) 
\frac{\chi^{2}}{\chi_{0}^{2}}
\nonumber \\
& + &  g_4 (\omega ^4 +6\omega^2 \rho^2+\rho^4 + 2\phi^4),
\end{eqnarray}
\begin{eqnarray}
{\cal L} _{0} & = & -\frac{1}{2} k_{0}\chi^{2} \left( \sigma^{2} + \zeta^{2} 
+ \delta^{2} \right) + k_{1} \left( \sigma^{2} + \zeta^{2} + \delta^{2} 
\right)^{2} \nonumber\\
&+& k_{2} \left( \frac{\sigma^{4}}{2} + \frac{\delta^{4}}{2} + 3 \sigma^{2} 
\delta^{2} + \zeta^{4} \right) 
+ k_{3}\chi\left( \sigma^{2} - \delta^{2} \right)\zeta \nonumber\\
&-& k_{4} \chi^{4} - \frac{1}{4} \chi^{4} {\rm {ln}} 
\frac{\chi^{4}}{\chi_{0}^{4}} 
+ \frac{d}{3} \chi^{4} {\rm {ln}} \Bigg (\bigg( \frac{\left( \sigma^{2} 
- \delta^{2}\right) \zeta }{\sigma_{0}^{2} \zeta_{0}} \bigg) 
\bigg (\frac{\chi}{\chi_0}\bigg)^3 \Bigg ),
\label{lagscal}
\end{eqnarray}
and 
\begin{eqnarray}
{\cal L} _{SB} & = & - \left( \frac{\chi}{\chi_{0}}\right) ^{2} 
\left[ m_{\pi}^{2} 
f_{\pi} \sigma + \left( \sqrt{2} m_{k}^{2}f_{k} - \frac{1}{\sqrt{2}} 
m_{\pi}^{2} f_{\pi} \right) \zeta \right]. 
\end{eqnarray}
In (\ref{lagscvec}), ${m_i}^*$ is the effective mass of the baryon of 
species $i$, given as
\begin{equation}
{m_i}^{*} = -(g_{\sigma i}\sigma + g_{\zeta i}\zeta + g_{\delta i}\delta)
\label{mbeff}
\end{equation}
The baryon-scalar meson interactions, as can be seen from equation
(\ref{mbeff}), generate the baryon masses through 
the coupling of  baryons to the non-strange $\sigma$, strange $\zeta$ 
scalar mesons and also to scalar-isovector meson $\delta$. In analogy 
to the baryon-scalar meson coupling there exist two independent 
baryon-vector meson interaction terms corresponding to the F-type 
(antisymmetric) and D-type (symmetric) couplings. Here antisymmetric 
coupling is used because the universality principle \cite{saku69} 
and vector meson dominance model suggest small symmetric coupling. 
Additionally,  we choose the parameters \cite{paper3,isoamss} so as 
to decouple the strange vector field $\phi_{\mu}\sim\bar{s}\gamma_{\mu}s$ 
from the nucleon, corresponding to an ideal mixing between $\omega$ and 
$\phi$ mesons. A small deviation of the mixing angle from ideal mixing 
\cite{dumbrajs,rijken,hohler1} has not been taken into account in the 
present investigation.

The concept of broken scale invariance leading to the trace anomaly 
in (massless) QCD, $\theta_{\mu}^{\mu} = \frac{\beta_{QCD}}{2g} 
G_{\mu\nu}^{a} G^{\mu\nu a}$, where $G_{\mu\nu}^{a} $ is the 
gluon field strength tensor of QCD, is simulated in the effective 
Lagrangian at tree level \cite{sche1} through the introduction of 
the scale breaking terms 
\begin{equation}
{\cal L}_{scalebreaking} =  -\frac{1}{4} \chi^{4} {\rm {ln}}
\Bigg ( \frac{\chi^{4}} {\chi_{0}^{4}} \Bigg ) + \frac{d}{3}{\chi ^4} 
{\rm {ln}} \Bigg ( \bigg (\frac{I_{3}}{{\rm {det}}\langle X 
\rangle _0} \bigg ) \bigg ( \frac {\chi}{\chi_0}\bigg)^3 \Bigg ),
\label{scalebreak}
\end{equation}
where $I_3={\rm {det}}\langle X \rangle$, with $X$ as the multiplet
for the scalar mesons. These scale breaking terms,
in the mean field approximation, are given by the last two terms
of the Lagrangian denstiy, ${\cal L}_0$  given by equation (\ref{lagscal}). 
The effect of these logarithmic terms is to break the scale invariance, 
which leads to the trace of the energy momentum tensor as \cite{heide1}
\begin{equation}
\theta_{\mu}^{\mu} = \chi \frac{\partial {\cal L}}{\partial \chi} 
- 4{\cal L} 
= -(1-d)\chi^{4}.
\label{tensor1}
\end{equation}
Hence the scalar gluon condensate of QCD ($\frac {\alpha_s}{\pi}
\langle {G^a}_{\mu \nu}
G^{\mu \nu a} \rangle$) is simulated by a scalar dilaton field in the present
hadronic model. 

The coupled equations of motion for the non-strange scalar field $\sigma$, 
strange scalar field $ \zeta$, scalar-isovector field $ \delta$ and dilaton 
field $\chi$, are derived from the Lagrangian density
and are given as
\begin{eqnarray}
&& k_{0}\chi^{2}\sigma-4k_{1}\left( \sigma^{2}+\zeta^{2}
+\delta^{2}\right)\sigma-2k_{2}\left( \sigma^{3}+3\sigma\delta^{2}\right)
-2k_{3}\chi\sigma\zeta \nonumber\\
&-&\frac{d}{3} \chi^{4} \bigg (\frac{2\sigma}{\sigma^{2}-\delta^{2}}\bigg )
+\left( \frac{\chi}{\chi_{0}}\right) ^{2}m_{\pi}^{2}f_{\pi}
-\sum g_{\sigma i}\rho_{i}^{s} = 0 
\label{sigma}
\end{eqnarray}
\begin{eqnarray}
&& k_{0}\chi^{2}\zeta-4k_{1}\left( \sigma^{2}+\zeta^{2}+\delta^{2}\right)
\zeta-4k_{2}\zeta^{3}-k_{3}\chi\left( \sigma^{2}-\delta^{2}\right)\nonumber\\
&-&\frac{d}{3}\frac{\chi^{4}}{\zeta}+\left(\frac{\chi}{\chi_{0}} \right) 
^{2}\left[ \sqrt{2}m_{k}^{2}f_{k}-\frac{1}{\sqrt{2}} m_{\pi}^{2}f_{\pi}\right]
 -\sum g_{\zeta i}\rho_{i}^{s} = 0 
\label{zeta}
\end{eqnarray}
\begin{eqnarray}
& & k_{0}\chi^{2}\delta-4k_{1}\left( \sigma^{2}+\zeta^{2}+\delta^{2}\right)
\delta-2k_{2}\left( \delta^{3}+3\sigma^{2}\delta\right) +k_{3}\chi\delta 
\zeta \nonumber\\
& + &  \frac{2}{3} d \chi^{4} \left( \frac{\delta}{\sigma^{2}-\delta^{2}}\right)
-\sum g_{\delta i}\rho_{i}^{s} = 0
\label{delta}
\end{eqnarray}
 
\begin{eqnarray}
& & k_{0}\chi \left( \sigma^{2}+\zeta^{2}+\delta^{2}\right)-k_{3}
\left( \sigma^{2}-\delta^{2}\right)\zeta + \chi^{3}\left[1
+{\rm {ln}}\left( \frac{\chi^{4}}{\chi_{0}^{4}}\right)  \right]
+(4k_{4}-d)\chi^{3}
\nonumber\\
& - & \frac{4}{3} d \chi^{3} {\rm {ln}} \Bigg ( \bigg (\frac{\left( \sigma^{2}
-\delta^{2}\right) \zeta}{\sigma_{0}^{2}\zeta_{0}} \bigg ) 
\bigg (\frac{\chi}{\chi_0}\bigg)^3 \Bigg ) 
+\frac{2\chi}{\chi_{0}^{2}}\left[ m_{\pi}^{2}
f_{\pi}\sigma +\left(\sqrt{2}m_{k}^{2}f_{k}-\frac{1}{\sqrt{2}}
m_{\pi}^{2}f_{\pi} \right) \zeta\right]  = 0 
\label{chi}
\end{eqnarray}
In the above, ${\rho_i}^s$ are the scalar densities for the baryons, 
given as 
\begin{eqnarray}
\rho_{i}^{s} = \gamma_{i}\int\frac{d^{3}k}{(2\pi)^{3}} 
\frac{m_{i}^{*}}{E_{i}^{*}(k)} 
\Bigg ( \frac {1}{e^{({E_i}^* (k) -{\mu_i}^*)/T}+1}
+ \frac {1}{e^{({E_i}^* (k) +{\mu_i}^*)/T}+1} \Bigg )
\label{scaldens}
\end{eqnarray}
where, ${E_i}^*(k)=(k^2+{{m_i}^*}^2)^{1/2}$, and, ${\mu _i}^* 
=\mu_i -g_{\omega i}\omega -g_{\rho i}\rho -g_{\phi i}\phi$, are the single 
particle energy and the effective chemical potential
for the baryon of species $i$, and,
$\gamma_i$=2 is the spin degeneracy factor \cite{isoamss}.

The above coupled equations of motion are solved to obtain the density 
and temperature dependent values of the scalar fields ($\sigma$,
$\zeta$ and $\delta$) and the dilaton field, $\chi$, in the isospin
asymmetric hot nuclear medium. As has already been mentioned, the value 
of the $\chi$ is related to the scalar gluon condensate in the hot 
hadronic medium, and is used to compute the in-medium masses of charmonium 
states in the present investigation. The isospin asymmetry in the medium
is introduced through the scalar-isovector field $\delta$ 
and therefore the dilaton field obtained after solving the above 
equations is also dependent on the isospin asymmetry parameter,
$\eta$ defined as $\eta= ({\rho_n -\rho_p})/({2 \rho_B})$, 
where $\rho_n$ and $\rho_p$ are the number densities of the neutron
and the proton and $\rho_B$ is the baryon density. In the present 
investigation, we study the effect of isospin asymmetry of the medium 
on the masses of the charmonium states $J/\psi, \psi(3686)$ 
and $\psi(3770)$.

The comparison of the trace of the energy momentum tensor arising
from the trace anomaly of QCD with that of the present chiral model
gives the relation of the dilaton field to the scalar gluon condensate.
We have, in the limit of massless quarks \cite{cohen},
\begin{equation}
\theta_{\mu}^{\mu} = \langle \frac{\beta_{QCD}}{2g} 
G_{\mu\nu}^{a} G^{\mu\nu a} \rangle  \equiv  -(1 - d)\chi^{4} 
\label{tensor2}
\end{equation}
The parameter $d$ originates from the second logarithmic term of equation 
(\ref{scalebreak}). To get an insight into the value of the parameter 
$d$, we recall that the QCD $\beta$ function at one loop level, for 
$N_{c}$ colors and $N_{f}$ flavors is given by
\begin{equation}
\beta_{\rm {QCD}} \left( g \right) = -\frac{11 N_{c} g^{3}}{48 \pi^{2}} 
\left( 1 - \frac{2 N_{f}}{11 N_{c}} \right)  +  O(g^{5})
\label{beta}
\end{equation}
In the above equation, the first term in the parentheses arises from 
the (antiscreening) self-interaction of the gluons and the second term, 
proportional to $N_{f}$, arises from the (screening) contribution of 
quark pairs. Equations (\ref{tensor2}) and (\ref{beta}) suggest the 
value of $d$ to be 6/33 for three flavors and three colors, and 
for the case of three colors and two flavors, the value of $d$ 
turns out to be 4/33, to be consistent with the one loop estimate 
of QCD $\beta$ function. These values give the order of magnitude 
about which the parameter $d$ can be taken \cite{heide1}, since one 
cannot rely on the one-loop estimate for $\beta_{\rm {QCD}}(g)$. 
In the present investigation of the in-medium properties of the 
charmonium states due to the medium modification of the dilaton 
field within chiral $SU(3)$ model, we use the value of $d$=0.064
\cite{amarind}. This parameter, along with the other parameters
corresponding to the  scalar Lagrangian density, ${\cal L}_0$ 
given by (\ref{lagscal}), are fitted so as to ensure 
extrema in the vacuum for the $\sigma$, $\zeta$ and $\chi$ field 
equations, to  reproduce the vacuum masses of the $\eta$ and $\eta '$ 
mesons, the mass of the $\sigma$ meson around 500 MeV, and pressure, p($\rho_0$)=0,
with $\rho_0$ as the nuclear matter saturation density \cite{paper3,amarind}.

The trace of the energy-momentum tensor in QCD, using the 
one loop beta function given by equation (\ref{beta}),
for $N_c$=3 and $N_f$=3, is given as,
\begin{equation}
\theta_{\mu}^{\mu} = - \frac{9}{8} \frac{\alpha_{s}}{\pi} 
G_{\mu\nu}^{a} G^{\mu\nu a}
\label{tensor4}
\end{equation} 
Using equations (\ref{tensor2}) and (\ref{tensor4}), we can write  
\begin{equation}
\left\langle  \frac{\alpha_{s}}{\pi} G_{\mu\nu}^{a} G^{\mu\nu a} 
\right\rangle =  \frac{8}{9}(1 - d) \chi^{4}
\label{chiglu}
\end{equation}
We thus see from the equation (\ref{chiglu}) that the scalar 
gluon condensate $\left\langle \frac{\alpha_{s}}{\pi} G_{\mu\nu}^{a} 
G^{\mu\nu a}\right\rangle$ is proportional to the fourth power of the 
dilaton field, $\chi$, in the chiral SU(3) model.
As mentioned earlier, the in-medium masses of charmonium states are 
modified due to the gluon condensates. Therefore, we need to know the 
change in the gluon condensate with density and temperature
of the asymmetric nuclear medium, which is calculated from the 
modification of the $\chi$ field, by using equation (\ref{chiglu}). 

\section{$D$ and $\bar D$ mesons in hot asymmetric nuclear matter}

In this section we study the $D$ and $\bar{D}$ mesons properties in 
isospin-asymmetric nuclear matter at finite temperatures. 
The medium modifications of the $D$ and $\bar D$ mesons arise due
to their interactions with the nucleons and the scalar mesons and the 
interaction Lagrangian density is given as \cite{amarind}
\begin{eqnarray}
\cal L _{DN} & = & -\frac {i}{8 f_D^2} \Big [3\Big (\bar p \gamma^\mu p
+\bar n \gamma ^\mu n \Big) 
\Big({D^0} (\partial_\mu \bar D^0) - (\partial_\mu {{D^0}}) {\bar D}^0 \Big )
+\Big(D^+ (\partial_\mu D^-) - (\partial_\mu {D^+})  D^- \Big )
\nonumber \\
& +&
\Big (\bar p \gamma^\mu p -\bar n \gamma ^\mu n \Big) 
\Big({D^0} (\partial_\mu \bar D^0) - (\partial_\mu {{D^0}}) {\bar D}^0 \Big )
- \Big( D^+ (\partial_\mu D^-) - (\partial_\mu {D^+})  D^- \Big )
\Big ]
\nonumber \\
 &+ & \frac{m_D^2}{2f_D} \Big [ 
(\sigma +\sqrt 2 \zeta_c)\big (\bar D^0 { D^0}+(D^- D^+) \big )
 +\delta \big (\bar D^0 { D^0})-(D^- D^+) \big )
\Big ] \nonumber \\
& - & \frac {1}{f_D}\Big [ 
(\sigma +\sqrt 2 \zeta_c )
\Big ((\partial _\mu {{\bar D}^0})(\partial ^\mu {D^0})
+(\partial _\mu {D^-})(\partial ^\mu {D^+}) \Big )
\nonumber \\
 & + & \delta
\Big ((\partial _\mu {{\bar D}^0})(\partial ^\mu {D^0})
-(\partial _\mu {D^-})(\partial ^\mu {D^+}) \Big )
\Big ]
\nonumber \\
&+ & \frac {d_1}{2 f_D^2}(\bar p p +\bar n n 
 )\big ( (\partial _\mu {D^-})(\partial ^\mu {D^+})
+(\partial _\mu {{\bar D}^0})(\partial ^\mu {D^0})
\big )
\nonumber \\
&+& \frac {d_2}{4 f_D^2} \Big [
(\bar p p+\bar n n))\big ( 
(\partial_\mu {\bar D}^0)(\partial^\mu {D^0})
+ (\partial_\mu D^-)(\partial^\mu D^+) \big )\nonumber \\
 &+&  (\bar p p -\bar n n) \big ( 
(\partial_\mu {\bar D}^0)(\partial^\mu {D^0})\big )
- (\partial_\mu D^-)(\partial^\mu D^+) ) 
\Big ]
\label{ldn}
\end{eqnarray}
In Eq.(\ref{ldn}), the first term is the vectorial Weinberg Tomozawa 
interaction term, obtained from the kinetic term of Eq.(1). The second 
term is obtained from 
the explicit symmetry breaking term and leads to the attractive interactions 
for both the $D$ and $\bar{D}$ mesons in the medium. The next three terms of 
above Lagrangian density ($\sim (\partial_\mu {\bar D})(\partial ^\mu D)$)
are known as the range terms. The first range term (with coefficient 
$\big (-\frac{1}{f_D}\big)$) is obtained from the kinetic energy term 
of the pseudoscalar mesons. The second and third range terms $d_{1}$ 
and $d_{2}$ are written for the $DN$ interactions in analogy with 
those written for $KN$ interactions in \cite{isoamss2}. It might be 
noted here that the interaction of the pseudoscalar mesons with the 
vector mesons, in addition to the pseudoscalar meson-nucleon vectorial 
interaction, leads to a double counting in the linear realization of 
chiral effective theories. Further, in the non-linear realization,
such an interaction does not arise in the leading or subleading order, 
but only as a higher order contribution \cite{borasoy}. Hence the 
vector meson-pseudoscalar interactions will not be taken into account
in the present investigation.

The  dispersion relations for the $D$ and $\bar{D}$ mesons are obtained 
by the Fourier transformations of equations of motion. These are given as 
\begin{equation}
-\omega^{2}+\vec{k}^{2}+m_{D}^{2}-\Pi\left(\omega,\vert\vec{k}\vert\right) 
= 0
\end{equation}
where, $m_D$ is the vacuum mass of the $D(\bar D)$ meson
taken as 1869 MeV and 1864.5 MeV for the $D$ and $\bar D$ mesons 
respectively. $\Pi\left(\omega,\vert\vec{k}\vert\right)$ denotes 
the self-energy of the $D\left( \bar{D} \right) $ mesons in the medium.

The self-energy $\Pi\left( \omega , \vert\vec{k}\vert\right) $ for the $D$ 
meson doublet $ \left( D^{0} , D^{+}\right) $ arising from the interaction 
of Eq.(\ref{ldn}) is given as
\begin{eqnarray}
\Pi (\omega, |\vec k|) &= & \frac {1}{4 f_D^2}\Big [3 (\rho_p +\rho_n)
\pm (\rho_p -\rho_n) 
\Big ] \omega \nonumber \\
&+&\frac {m_D^2}{2 f_D} (\sigma ' +\sqrt 2 {\zeta_c} ' \pm \delta ')
\nonumber \\ & +& \Big [- \frac {1}{f_D}
(\sigma ' +\sqrt 2 {\zeta_c} ' \pm \delta ')
+\frac {d_1}{2 f_D ^2} (\rho_s ^p +\rho_s ^n)\nonumber \\
&+&\frac {d_2}{4 f_D ^2} \Big (({\rho^s} _p +{\rho^s} _n)
\pm   ({\rho^s} _p -{\rho^s} _n) \Big ) \Big ]
(\omega ^2 - {\vec k}^2),
\label{selfd}
\end{eqnarray}

where the $\pm$ signs refer to the $D^{0}$ and $D^{+}$ mesons, 
respectively, and $\sigma^{\prime}\left( \sigma - \sigma_{0}\right) $, 
$\zeta_{c}^{\prime}\left(\zeta_{c} - \zeta_{c0}\right)$, and 
$\delta^{\prime}\left(  = \delta -\delta_{0}\right) $ are the 
fluctuations of the scalar isoscalar fields $\sigma$ and $\zeta$ and
the scalar-isoscalar field $\delta$ from their 
vacuum expectation values. The vacuum expectation value of $\delta$ 
is zero $\left(\delta_{0}=0 \right)$, since a nonzero value for it 
will break the isospin-symmetry of the vacuum. (We neglect here the 
small isospin breaking effect arising  from the mass and charge 
difference of the up and down quarks.) 
We might note here that the interaction of the scalar quark condensate 
$\zeta_{c}$ (being made up of heavy charmed quarks and antiquarks)
leads to very small modifications of the masses \cite{roeder}. So we will 
not consider the medium fluctuations of $\zeta_{c}$. 
In the present investigation, we take the value of the D meson
decay constant, $f_D$ as 135 MeV \cite{weise}. Within the present
model, the medium modification to the $D(\bar D)$ mesons due to
the scalar interaction depends only on the fluctuations of the scalar 
$\sigma$ and $\delta$ fields in the asymmetric hot nuclear medium  
which are determined by solving the coupled equations (\ref{sigma}),
(\ref{zeta}), (\ref{delta}) and (\ref{chi}). As the scalar term
of (\ref{selfd}) does not contain any free parameters, with the
assumption that the fluctuation of the charm condensate has 
a negligible effect on the masses of the $D(\bar D)$ mesons,
there are no uncertainties in the mass shift due to the scalar interaction
in the present investigation, once the value for $f_D$ is chosen.
In Eq. (\ref{selfd}),
$\rho_{p}$ and $\rho_{n}$ are the number densities of protons 
and neutrons given by
\begin{equation}
\rho_{i} = \gamma_{i} \int \frac{d^{3}k}{(2\pi)^{3}} 
 \left( \frac{1}{e^{\left( E_{i}^{*}(k) - \mu_{i}^{*}\right) /T} + 1} 
-  \frac{1}{e^{\left( E_{i}^{*}(k) + \mu_{i}^{*}\right) /T} + 1}\right), 
\label{vecdens}
\end{equation}
for $i$=p and n, and $\rho_{p}^{s}$ and $\rho_{n}^{s}$ are their 
scalar densities, as given by equation (\ref{scaldens}).
  
Similarly, for the  $\bar{D}$ meson doublet 
$\left(\bar{D}^{0},D^{-}\right)$, the self-energy is calculated as 
\begin{eqnarray}
\Pi (\omega, |\vec k|) &= & -\frac {1}{4 f_D^2}\Big [3 (\rho_p +\rho_n)
\pm (\rho_p -\rho_n) \Big ] \omega\nonumber \\
&+&\frac {m_D^2}{2 f_D} (\sigma ' +\sqrt 2 {\zeta_c} ' \pm \delta ')
\nonumber \\ & +& \Big [- \frac {1}{f_D}
(\sigma ' +\sqrt 2 {\zeta_c} ' \pm \delta ')
+\frac {d_1}{2 f_D ^2} (\rho_s ^p +\rho_s ^n
)\nonumber \\
&+&\frac {d_2}{4 f_D ^2} \Big (({\rho^s} _p +{\rho^s} _n)
\pm   ({\rho^s} _p -{\rho^s} _n) \Big ]
(\omega ^2 - {\vec k}^2),
\label{selfdbar}
\end{eqnarray}
where the $\pm$ signs refer to the $\bar{D}^{0}$ and $D^{-}$ mesons, 
respectively. The optical potentials of the $D$ and $\bar{D}$ mesons are 
obtained using the expression
\begin{equation}
U(\omega, k) = \omega(k) - \sqrt{k^{2} + m_{D}^{2}}
\end{equation}
where $m_{D}$ is the vacuum mass for the $D(\bar{D})$ meson and 
$\omega(k)$ is the momentum-dependent energy of the $D(\bar{D})$ meson.

\section{Charmonium masses in hot asymmetric nuclear matter}
In this section, we investigate the masses of charmonium states $J/\psi$, 
$\psi(3686)$ and $\psi(3770)$, in isospin asymmetric hot nuclear matter. 
From the QCD sum rule calculations, the mass shift of the charmonium 
states in the medium is due to the gluon condensates \cite{leeko,arata}.
For heavy quark systems, there are two independent lowest dimension
operators: the scalar gluon condensate ($\left\langle 
\frac{\alpha_{s}}{\pi} G_{\mu\nu}^{a} G^{\mu\nu a}\right\rangle$) 
and the condensate of the twist 2 gluon operator ($\left\langle 
\frac{\alpha_{s}}{\pi} G_{\mu\nu}^{a} G^{\mu\alpha a}\right\rangle$). 
These operators can be rewritten in terms of the color electric and 
color magnetic fields, $\langle \frac{\alpha_s}{\pi} {\vec E}^2\rangle$ 
and $\langle \frac{\alpha_s}{\pi} {\vec B}^2\rangle$. Additionally, 
since the Wilson coefficients for the operator $\langle \frac{\alpha_s}{\pi} 
{\vec B}^2\rangle$ vanish in the non-relativistic limit, the only 
contribution from the gluon condensates is proportional to $\langle 
\frac{\alpha_s}{\pi} {\vec E}^2\rangle$, similar to the second order 
Stark effect.  Hence, the mass shift of the charmonium states  arises
due to the change in the operator $\langle \frac{\alpha_s}{\pi} 
{\vec E}^2\rangle$ in the medium from its vacuum value \cite {leeko}. 
In the leading order mass shift formula derived in the large charm 
mass limit \cite{pes1}, the shift in the mass of the charmonium 
state is given as
\cite{leeko}
\begin{equation}
\Delta m_{\psi} (\epsilon) = -\frac{1}{9} \int dk^{2} \vert 
\frac{\partial \psi (k)}{\partial k} \vert^{2} \frac{k}{k^{2} 
/ m_{c} + \epsilon} \bigg ( 
\left\langle  \frac{\alpha_{s}}{\pi} E^{2} \right\rangle-
\left\langle  \frac{\alpha_{s}}{\pi} E^{2} \right\rangle_{0}
\bigg ).
\label{mass1}
\end{equation}
In the above, $m_c$ is the mass of the charm quark, taken as 1.95 GeV 
\cite{leeko}, $m_\psi$ is the vacuum mass of the charmonium state 
and $\epsilon = 2 m_{c} - m_{\psi}$. 
$\psi (k)$ is the wave function of the charmonium state
in the momentum space, normalized as $\int\frac{d^{3}k}{2\pi^{3}} 
\vert \psi(k) \vert^{2} = 1 $ \cite{leetemp}.
At finite densities, in the linear density approximation, the change 
in the value of $\langle \frac{\alpha_s}{\pi} {\vec E}^2\rangle$, 
from its vacuum value, is given as 
\begin{equation}
\left\langle  \frac{\alpha_{s}}{\pi} E^{2} \right\rangle-
\left\langle  \frac{\alpha_{s}}{\pi} E^{2} \right\rangle_{0}
=
\left\langle  \frac{\alpha_{s}}{\pi} E^{2} \right\rangle _{N}
\frac {\rho_B}{2 M_N},
\end{equation}
and the mass shift in the charmonium states reduces to \cite{leeko}
\begin{equation}
\Delta m_{\psi} (\epsilon) = -\frac{1}{9} \int dk^{2} \vert 
\frac{\partial \psi (k)}{\partial k} \vert^{2} \frac{k}{k^{2} 
/ m_{c} + \epsilon} 
\left\langle  \frac{\alpha_{s}}{\pi} E^{2} \right\rangle _{N}
\frac {\rho_B}{2 M_N}.
\label{masslindens}
\end{equation}
In the above, $\left\langle  \frac{\alpha_{s}}{\pi} E^{2} 
\right\rangle _{N}$ is the expectation value of  
$\left\langle  \frac{\alpha_{s}}{\pi} E^{2} \right\rangle$
with respect to the nucleon.

The expectation value of the scalar gluon condensate can be expressed 
in terms of the color electric field and the color magnetic field 
as \cite{david}
\begin{equation}
\left\langle 
\frac{\alpha_{s}}{\pi} G_{\mu\nu}^{a} G^{\mu\nu a}\right\rangle 
=-2 \left\langle \frac{\alpha_{s}}{\pi} (E^{2} - B^{2}) \right\rangle.
\end{equation}
In the non-relativistic limit, as already mentioned, the contribution
from the magnetic field vanishes and hence, we can write,
\begin{equation}
\left\langle \frac{\alpha_{s}}{\pi} E^{2} \right\rangle
=-\frac {1}{2} 
\left\langle \frac{\alpha_{s}}{\pi} 
G_{\mu\nu}^{a} G^{\mu\nu a}\right\rangle 
\label{e2glu}
\end{equation}

Using equations (\ref{chiglu}), (\ref{mass1}) and (\ref{e2glu}), 
we obtain the expression for the mass shift in the charmonium 
in the hot and dense nuclear medium, which arises from 
the change in the dilaton field in the present investigation, 
as
\begin{equation}
\Delta m_{\psi} (\epsilon) = \frac{4}{81} (1 - d) \int dk^{2} 
\vert \frac{\partial \psi (k)}{\partial k} \vert^{2} \frac{k}{k^{2} 
/ m_{c} + \epsilon}  \left( \chi^{4} - {\chi_0}^{4}\right). 
\label{masspsi}
\end{equation}
In the above, $\chi$ and $\chi_0$ are the values of the dilaton field
in the nuclear medium and in the vacuum respectively.

In the present investigation, the wave functions for the charmonium states 
are taken to be Gaussian and are given as \cite{friman}
\begin{equation}
\psi_{N, l} = {\rm { Normalization}} \times Y_{l}^{m} (\theta, \phi) 
(\beta^{2} r^{2})^{\frac{1}2{} l} exp^{-\frac{1}{2} \beta^{2} r^{2}} 
L_{N - 1}^{l + \frac{1}{2}} \left( \beta^{2} r^{2}\right)
\label{wavefn} 
\end{equation} 
where $\beta^{2} = M \omega / h$ characterizes the strength of the 
harmonic potential, $M = m_{c}/2$ is the reduced mass of 
the charm quark and charm anti-quark system, and $L_{p}^{k} (z)$ 
is the associated Laguerre Polynomial. As in Ref. \cite{leeko},
the oscillator constant $\beta$ is determined from the mean squared 
radii $\langle r^{2} \rangle$ as 0.46$^{2}$ fm$^2$, 0.96$^{2}$ fm$^2$ 
and 1 fm$^{2}$ for the charmonium states $J/\psi(3097) $, $\psi(3686)$ and 
$\psi(3770)$, respectively. This gives the value for the parameter
$\beta$ as 0.51 GeV, 0.38 GeV and 0.37 GeV for $J/\psi(3097)$, 
$\psi(3686$ and $\psi(3770)$, assuming that these 
charmonium states are in the 1S, 2S and 1D states respectively. 
Knowing the wave functions of the charmonium states and 
calculating the medium modification of the dilaton field
in the hot nuclear matter, we obtain the mass shift of the
charmonium states, $J/\psi$, $\psi (3686)$ and $\psi (3770)$
respectively. In the next section we shall present the results 
of the present investigation of these in-medium charmonium masses 
in hot asymmetric nuclear matter.

\section{Results and Discussions}
\label{results}
In this section, we present the results of our investigation 
for the in-medium masses of $D$ and $\bar{D}$ mesons as well as 
of the charmonium states $J/\psi(3097)$, $\psi(3686)$ and $\psi(3770)$,
in isospin asymmetric nuclear matter at finite temperatures. We have 
generalized the chiral $SU(3)$ model to $SU(4)$ to include the 
interactions of the charmed mesons.  The values of the parameters used 
in the present investigation, are : $k_{0} = 2.54, k_{1} = 1.35, 
k_{2} = $-4.78$, k_{3} = -2.77$, $k_{4} = -0.22$ and $d =  0.064$, 
which are the parameters occurring in the scalar meson interactions 
defined in equation (\ref{lagscal}). The vacuum values of the scalar 
isoscalar fields, $\sigma$ and $\zeta$ and the dilaton field $\chi$ 
are $-93.3$ MeV, $-106.6$ MeV and $409.8$ MeV respectively. The values, 
$g_{\sigma N} = 10.6$ and $g_{\zeta N} = -0.47$ are determined by fitting 
to vacuum baryon masses. The other parameters fitted to the asymmetric 
nuclear matter saturation properties in the mean-field approximation 
are: $g_{\omega N}$ = 13.3, $g_{\rho p}$ = 5.5, $g_{4}$ = 79.7, 
$g_{\delta p}$ = 2.5, $m_{\zeta}$ = 1024.5 MeV, $ m_{\sigma}$ = 466.5 MeV 
and $m_{\delta}$ = 899.5 MeV. The nuclear matter saturation density 
used in the present investigation is $0.15$ fm$^{-3}$. The coefficients 
$d_{1}$ and $d_{2}$, calculated from the empirical values of the $KN$ 
scattering lengths for $I = 0$ and $I = 1$ channels, are $2.56/m_{K}$
and $0.73/m_{K}$, respectively \cite{isoamss2}. 
 
\begin{figure}
\includegraphics[width=16cm,height=16cm]{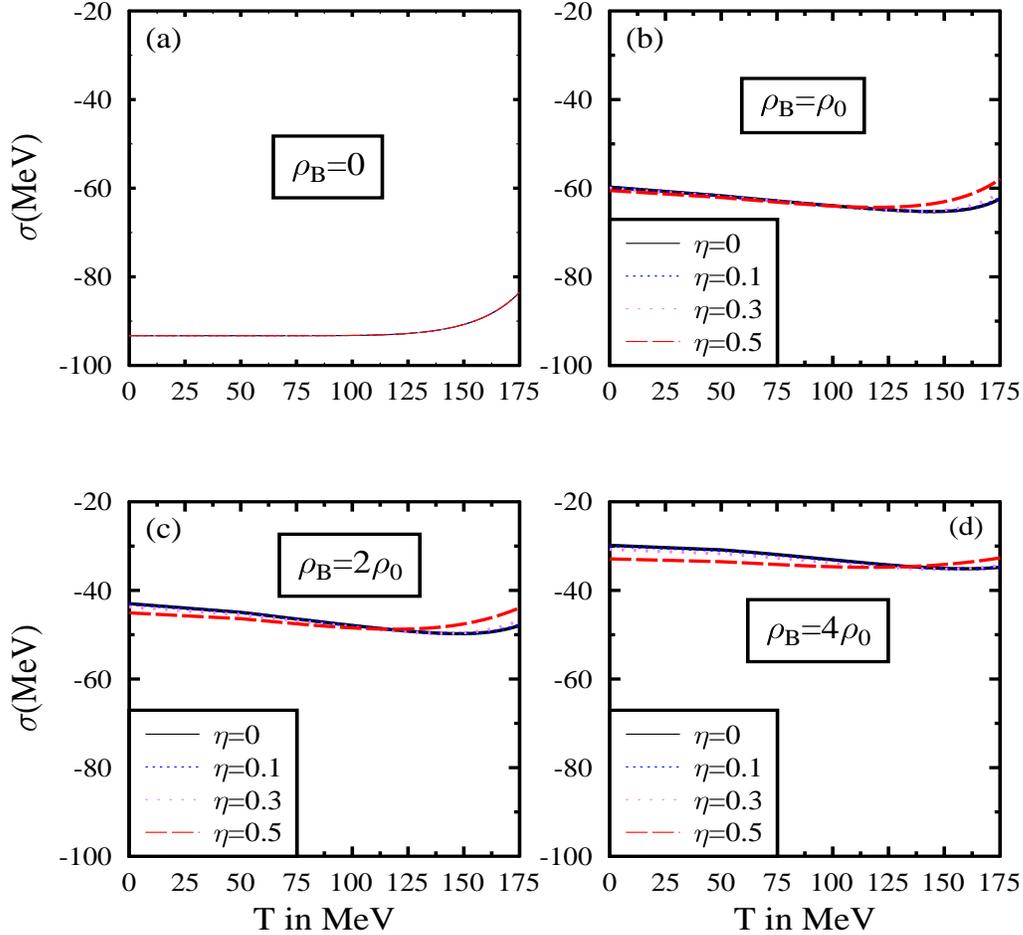}  
\caption{(Color online)
The scalar-isoscalar field $\sigma$ plotted as a function of temperature at a given baryon density ($\rho_{B} = 0, \rho_{0}, 2\rho_{0}$ and $4\rho_{0}$), for different values
of the isospin asymmetry parameter, $\eta$.
}
\label{fig1}
\end{figure}

\begin{figure}
\includegraphics[width=16cm,height=16cm]{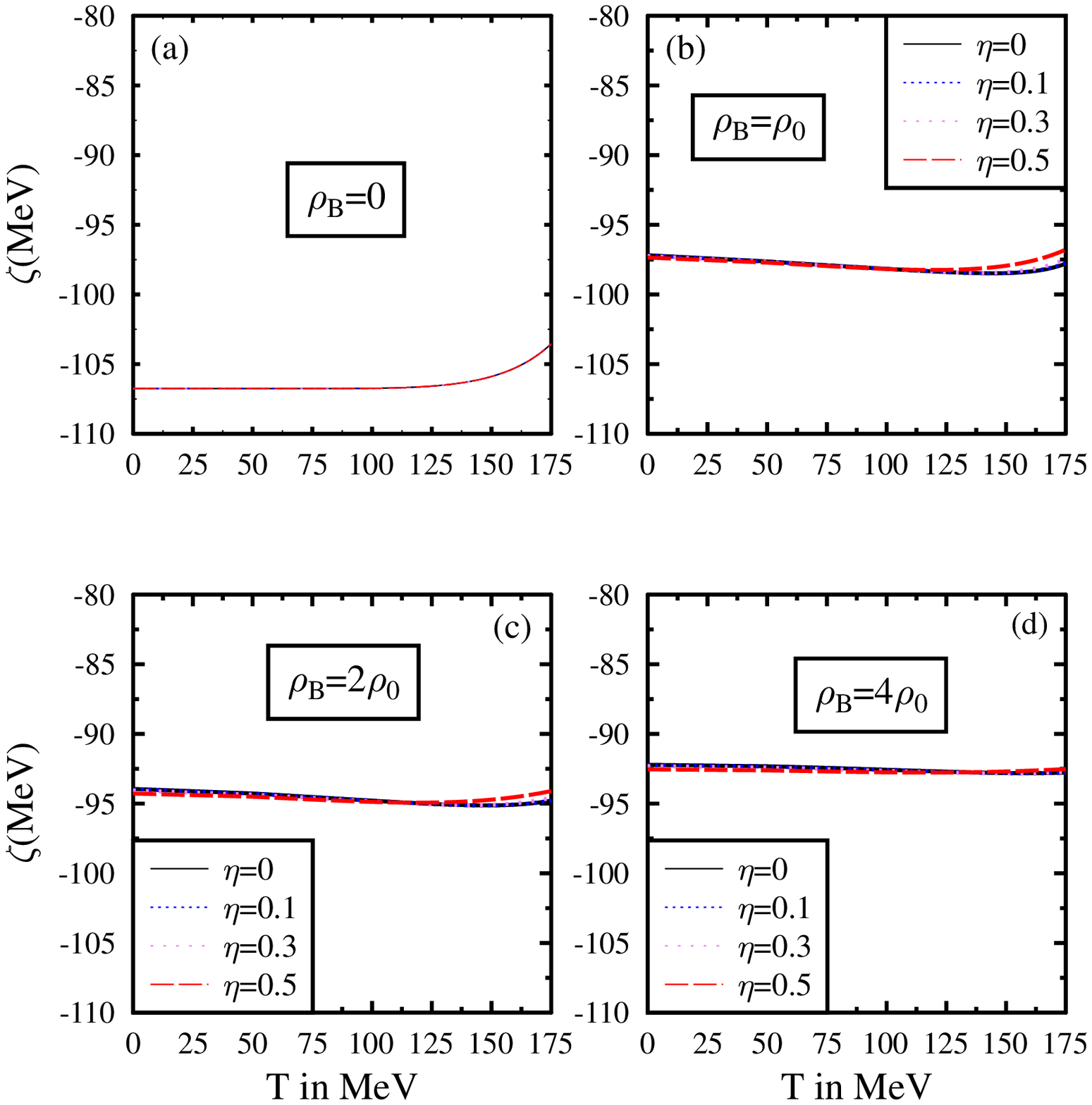}  
\caption{(Color online)
The scalar-isoscalar field $\zeta$ plotted as a function of temperature at a given baryon density ($\rho_{B} = 0, \rho_{0}, 2\rho_{0}$ and $4\rho_{0}$), for different values
of the isospin asymmetry parameter, $\eta$.
}
\label{fig2}
\end{figure}

\begin{figure}
\includegraphics[width=16cm,height=16cm]{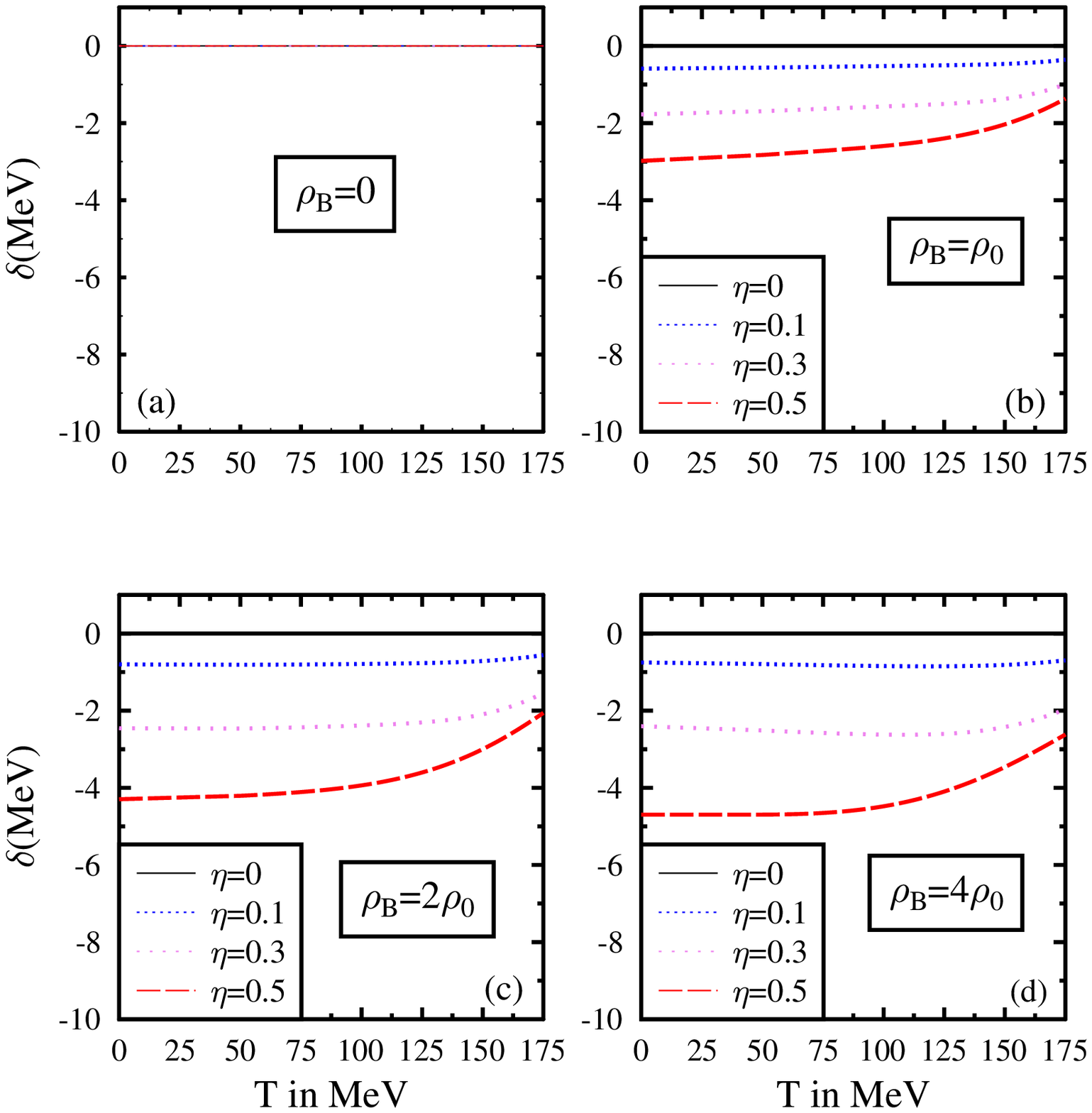}  
\caption{(Color online)
The scalar-isovector field $\delta$ plotted as a function of temperature at a given baryon density ($\rho_{B} = 0, \rho_{0}, 2\rho_{0}$ and $4\rho_{0}$), for different values
of the isospin asymmetry parameter, $\eta$.
}
\label{fig3}
\end{figure}

\begin{figure}
\includegraphics[width=16cm,height=16cm]{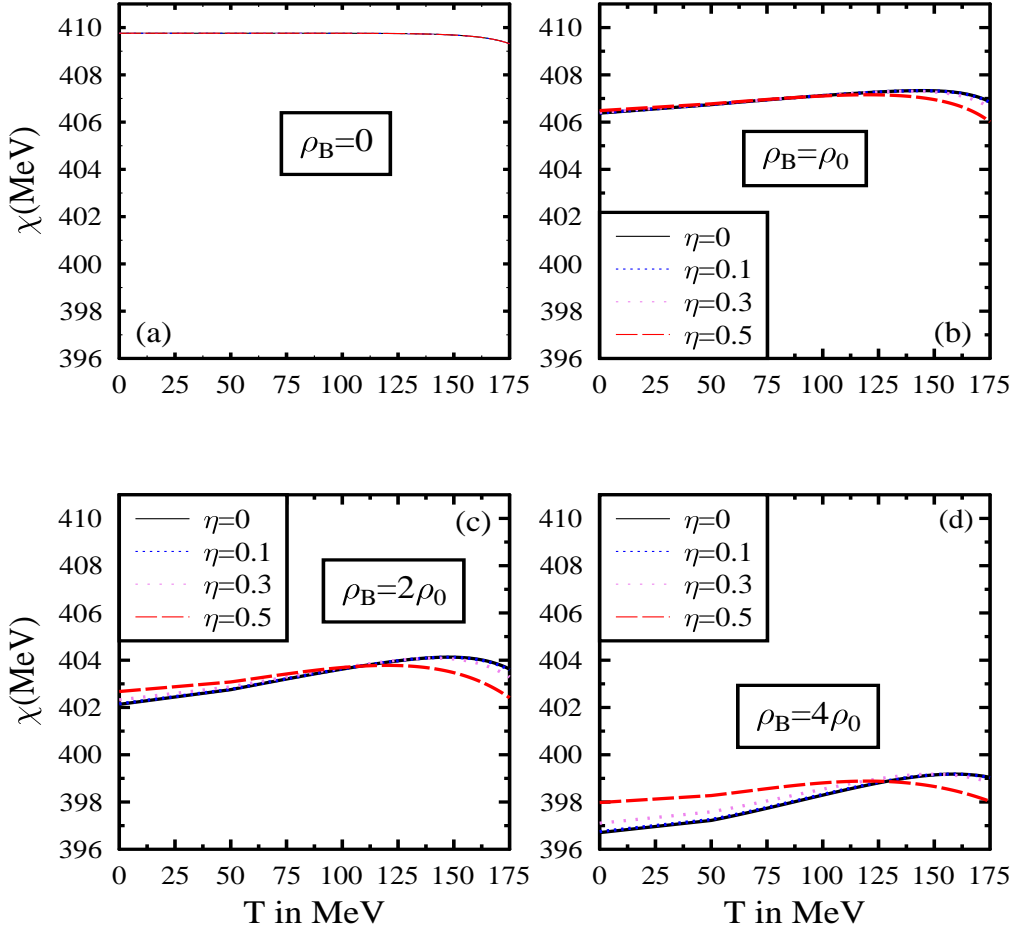}  
\caption{(Color online)
The dilaton field $\chi$ plotted as a function of temperature 
at a given baryon density ($\rho_{B} = 0, \rho_{0}, 2\rho_{0}$ 
and $4 \rho_{0}$), for different values of the isospin asymmetry 
parameter, $\eta$.
}
\label{fig4}
\end{figure}

In isospin asymmetric nuclear medium, the properties of the $D$ mesons 
($D^{0}$, $D^{+}$) and $\bar{D}$ mesons ($\bar{D}^{0}$,$D^{-}$), 
due to their interactions with the hot hadronic medium, undergo 
medium modifications. These modifications arise due to the interactions
with the nucleons (through the Weinberg-Tomozawa vectorial interaction
as well as through the range terms) and the scalar exchange terms.
The modifications of the scalar mean fields modify the masses of the
nucleons in the hot and dense hadronic medium. Before going into the 
details of how the $D$ and $\bar{D}$ mesons properties are modified 
at finite temperatures in the dense nuclear medium, let us see how 
the scalar fields are modified at finite temperatures in the nuclear 
medium. In figures \ref{fig1}, \ref{fig2}, \ref{fig3} and 
\ref{fig4}, we show the variation of scalar fields $\sigma$, 
$\zeta$ and scalar-isovector field $\delta$ and the dilaton field 
$\chi$, with temperature, for both zero and finite densities, and 
for selected values of the isospin asymmetry parameter, 
$\eta = 0, 0.1, 0.3$ and $0.5$.  At zero baryon density, we observe 
that the magnitudes of the scalar fields $\sigma$ and $\zeta$ 
decrease with increase in temperature. However, the drop in their magnitudes
with temperature is negligible upto a temperature of about 100 MeV
(that is, they remain very close to their vacuum values).
The changes in the magnitudes of the $\sigma$ and $\zeta$ 
fields are 2.5 MeV and 0.8 MeV respectively, when the temperature
changes from 100 MeV to 150 MeV, above which the drop increases.
These values change to about 10 MeV and 3 MeV for the $\sigma$
and $\zeta$ fields respectively, for a change in temperature of 
100 MeV to 175 MeV. At zero baryon density, it is observed that 
the value of the dilaton field remains almost constant
upto a temperature of about 130 MeV above which it is seen 
to drop with increase in temperature. However, the drop in 
the dilaton field is seen to be very small. The value of the 
dilaton field is seen to change from 409.8 MeV at T = 0 to 
about 409.7 MeV and 409.3 MeV at T = 150 MeV and T = 175 MeV, respectively. 
The thermal distribution functions have an effect of increasing the 
scalar densities at zero baryon density, i.e. $\mu_{i}^{*}=0$, as can be 
seen from the expression of the scalar densities, given by equation 
(\ref{scaldens}). This effect seems to be negligible upto a temperature 
of about 130 MeV. This leads to a decrease in the magnitudes of the
scalar fields, $\sigma$ and $\zeta$. This behaviour of the scalar fields 
is reflected in the value of $\chi$, which is solved from the coupled 
equations of motion of the scalar fields, given by equations (\ref{sigma}), 
(\ref{zeta}), (\ref{delta}) and (\ref{chi}), as a drop as we increase 
the temperature above a temperature of about 130 MeV.
The scalar densities attaining nonzero values at high temperatures, 
even at zero baryon density, indicates the presence of baryon-antibaryon 
pairs in the thermal bath and has already been observed in literature 
\cite{kristof1, frunstl1}. This leads to the baryon masses to be different 
from their vacuum masses above this temperature, arising from modifications 
of the scalar fields $\sigma$ and $\zeta$.

For finite density situation, the behaviour of the scalar fields with 
temperature is seen to be very different from the zero density case, 
as can be seen from the subplots (b), (c) and (d) of figures \ref{fig1}, 
\ref{fig2} and \ref{fig4} for $\sigma$, $\zeta$ and $\chi$ fields 
respectively, where the fields are plotted as functions of temperature 
for densities $\rho_{0}$, $2\rho_{0}$ and $4\rho_{0}$ respectively. 
At finite densities, one observes first a rise and then a decrease 
of the scalar fields $\sigma$, $\zeta$ and $\chi$ with temperature. 
For example, at $\rho_{B} = \rho_{0}$, and for the value of the 
isospin asymmetry parameter, $\eta =  0$ the scalar fields $\sigma$, 
$\zeta$ and $\chi$ increase with temperature upto a temperature 
of about 145 MeV and above this temperature, they both start decreasing. 
At $\eta =  0.5$ the value of temperature upto which these scalar fields 
increase becomes about 120 MeV. For $\rho_{B} = 4\rho_{0}$ and $\eta = 0$ 
the scalar fields $\sigma$, $\zeta$ and $\chi$ increase upto a temperature 
of about 160 MeV.
At $\eta = 0.5$ this value of temperature is lowered to about 120 MeV 
for $\sigma$, $\zeta$ and $\chi$ fields respectively. 
This observed rise in the magnitudes of $\sigma$ 
and $\zeta$ fields with temperature leads to an increase in the mass of 
nucleons with temperature at finite densities.
The reason for the different behaviour of the scalar fields 
($\sigma$ and $\zeta$) at zero and finite densities
can be understood in the following manner \cite{kristof1}. As has
already been mentioned, the thermal distribution functions in 
(\ref{scaldens}) have an effect of increasing the scalar densities 
at zero baryon density, i.e., for $\mu_{i}^{*} = 0$. However, 
at finite densities, i.e. for nonzero values of the effective 
chemical potential, $\mu_{i}^{*}$, for increasing temperature, 
there are contributions also from higher momenta, thereby, increasing 
the denominator of the integrand on the right hand side of equation 
(\ref{scaldens}). This leads to a decrease in the scalar density. 
The competing effects of the thermal distribution functions and 
the contributions from the higher moments states give rise to observed 
behaviour of the scalar density and hence of the $\sigma$ and $\zeta$ fields 
with temperature at finite baryon densities \cite{kristof1}. This kind 
of behaviour of the scalar $\sigma$ field on temperature at finite 
densities has also been observed in the Walecka model by Li and Ko 
\cite{liko}. The behaviour of the scalar fields $\sigma$ and $\zeta$ 
with the temperature is reflected in the behaviour of $\chi$ field, 
since it is solved from the coupled equations of the scalar fields. 
In figure \ref{fig3}, we observe that the value of the scalar 
isovector field, $\delta$ is zero at zero baryon density, since 
there is no isospin asymmetry at zero density. At finite 
baryon densities the magnitude of the scalar isovector field, 
$\delta$ decreases with increase in the temperature of the 
nuclear medium. However, for given temperature as we move 
to higher densities the magnitude of the $\delta$ field 
is seen to increase which means that the medium is more 
asymmetric at higher densities. 
In the isospin asymmetric nuclear medium, the scalar isovector
$\delta$ meson attains a nonzero expectation value. However,
the magnitudes of the $\delta$ field at a given baryon density
and temperature of the medium, which are solved from the coupled 
equations, (\ref{sigma}) to (\ref{chi}), for the scalar fields, 
turn out to be small ($\le 4-5 MeV$),as can be seen from figure 
\ref{fig3}. This causes changes in the $\sigma$ and $\zeta$ fields,
as can be seen from figures \ref{fig1} and \ref{fig2},
also to be small in the isospin asymmetric nuclear matter, 
since these solved from the equations (\ref{sigma}) and (\ref{zeta}), 
along with the equations for $\delta$ and $\chi$, given by
(\ref{delta}) and (\ref{chi}). 

\begin{figure}
\includegraphics[width=16cm,height=16cm]{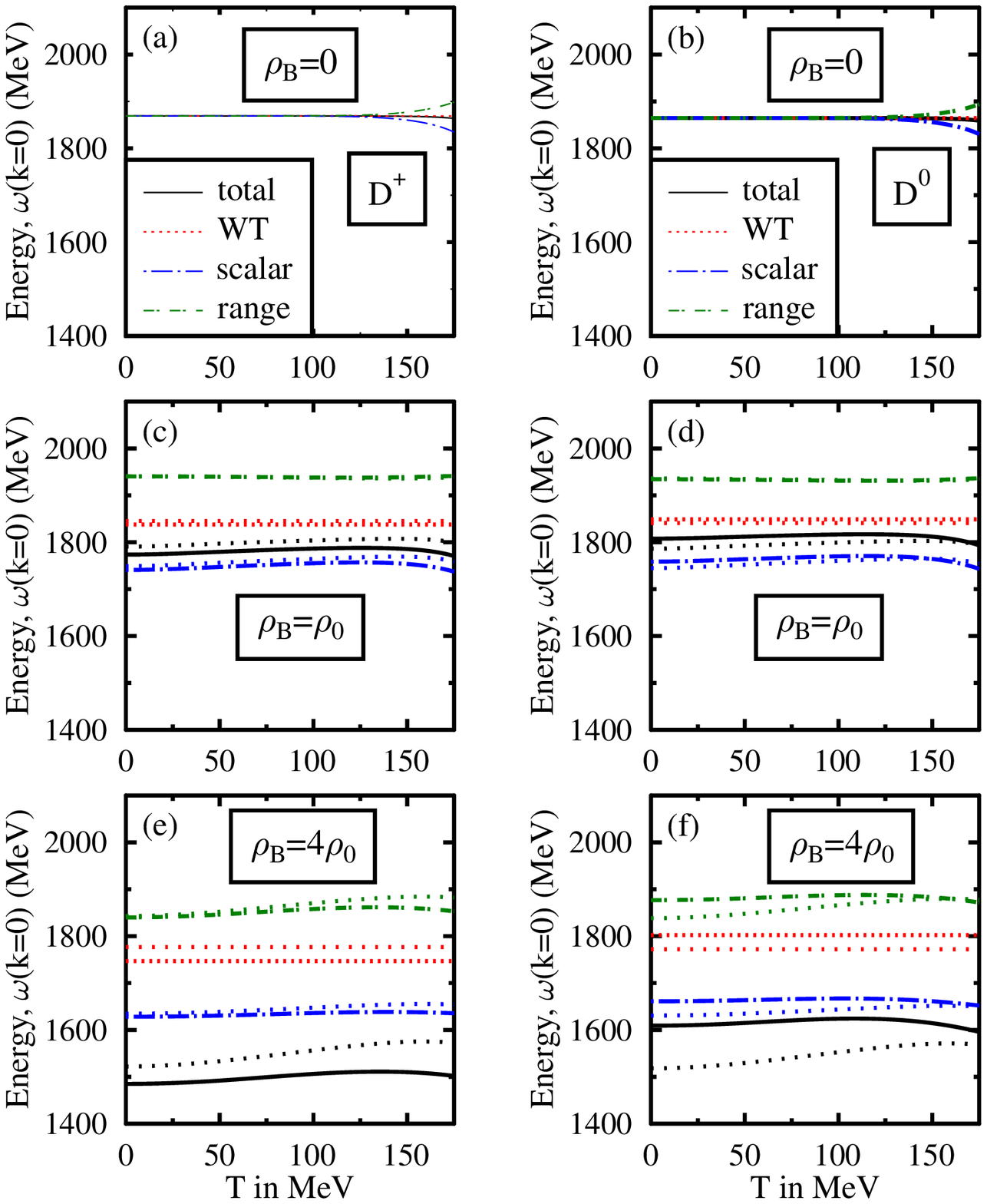} 
\caption{(Color online) The energies of $D^{+}$ meson ((a),(c) and (e)) 
and of $D^{0}$ meson ((b),(d) and (f)), at momentum $k = 0$, versus 
temperature, T, 
for different values of the isospin asymmetry 
parameter ($\eta = 0$ and $0.5$) and for given values of density 
($\rho_{B} = 0, \rho_{0}$ and $4\rho_{0}$). The values of parameters $d_{1}$ and $d_{2}$ are calculated from $KN$ scattering lengths
in I = 0 and I = 1 channels.} 
\label{fig5}
\end{figure}

\begin{figure}
\includegraphics[width=16cm,height=16cm]{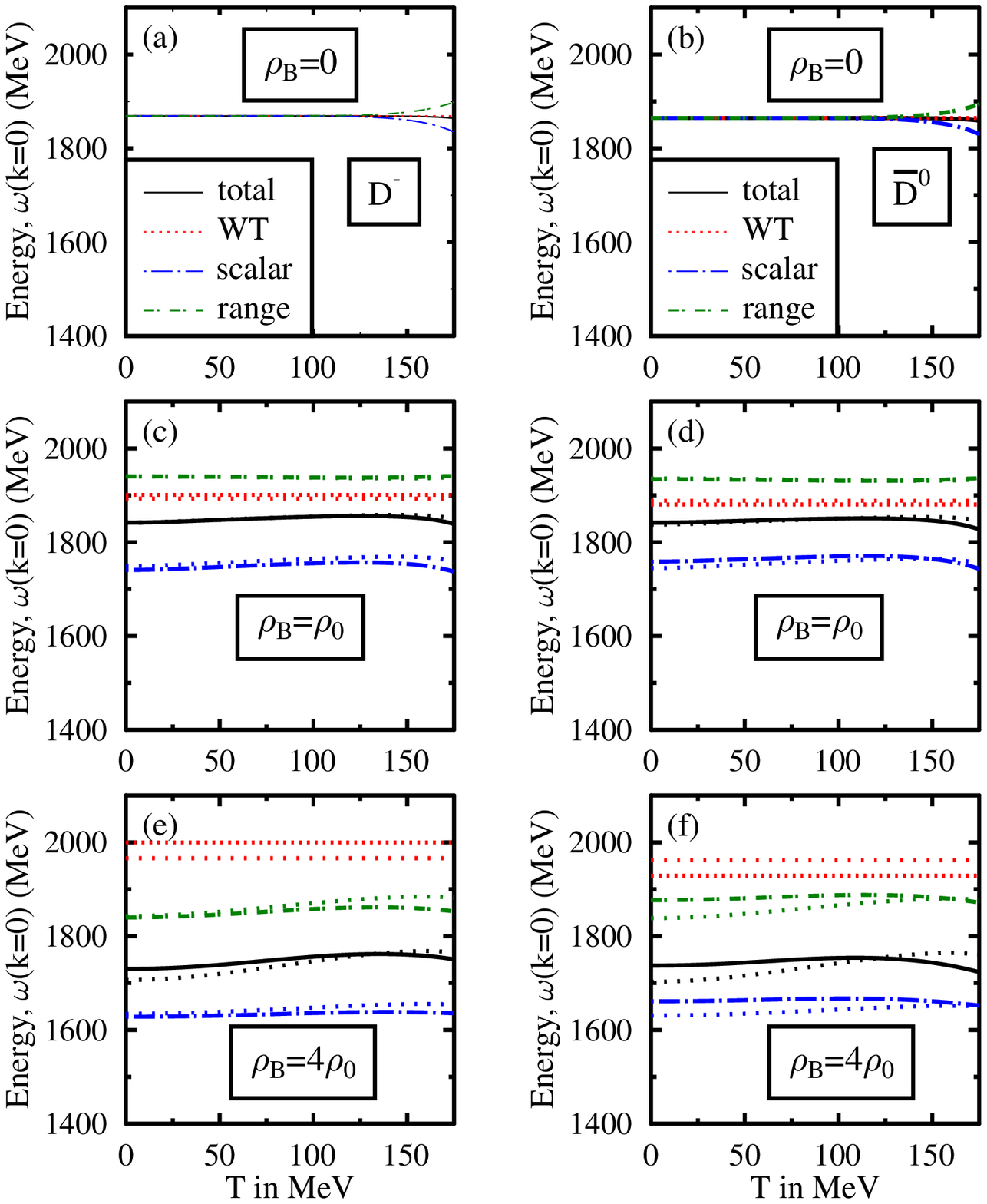} 
\caption{(Color online) The energies of $D^{-}$ meson ((a),(c) and (e)) 
and of $\bar{D^{0}}$ meson ((b),(d) and (f)), at momentum $k = 0$, versus 
temperature, T, 
, for different values of the isospin asymmetry 
parameter ($\eta = 0$ and $0.5$) and for given values of density 
($\rho_{B} = 0, \rho_{0}$ and $4\rho_{0}$). The values of parameters $d_{1}$ and $d_{2}$ are calculated from $KN$ scattering lengths
in I = 0 and I = 1 channels.} 
\label{fig6}
\end{figure}

In figures \ref{fig5} and \ref{fig6}, we show the variation of 
the energy of $D$ and $\bar{D}$ mesons due to Weinberg-Tomozawa term, 
the scalar exchange term arising from the explicit symmtery breaking term 
and the range terms, as functions of temperature and for the densities 
$\rho_{B} = 0, \rho_{0}$ and $4\rho_{0}$. These are plotted asymmetric
nuclear matter with
the value of the isospin asymmetry parameter as 0.5, and in the same
figure, the results arising from these contributions for
the isospin symmetric case ($\eta$=0) are shown as the dotted lines.
At zero density, the contributions from the Weinberg-Tomozawa term is zero,
since the densities of the proton and neutron,$\rho_p$ and $\rho_n$ 
are zero at zero density. The scalar fields attaining values different 
from their vacuum values at zero density at values of temperature higher 
than about 100 MeV, as already observed in figures \ref{fig1} and \ref{fig2},
leads to a decrease in the $D^+$ and $D^0$ mesons due to the
scalar meson exchange term and an increase in the contribution
due to the range term, as can be seen in the subplots (a) and (b) 
of figure \ref{fig5}. These two terms seem to almost cancel with each other,
resulting in negligible modifications of $D^+$ and $D^0$ mesons.
The drop in the masses of $D^+$ and $D^0$ are seen to be about 0.75 MeV and
0.75 MeV respectively at temperature T=150 MeV and about 5.4 MeV and 
5.35 MeV at a temperature of 175 MeV.
At finite densities, the contribution to the energies of the
$D^+$ and $D^0$ mesons arising from the Weinberg-Tomozawa term 
as a function of temperature, is constant, since this term depends 
on the densities of proton and neutron only, which for a given baryon 
density and a given value of the isospin asymmetry parameter,
are constant. At $\rho_B=\rho_0$ and T=0, the Weinberg-Tomozawa term gives 
a drop of about 23 MeV  in the masses of $D^+$ and $D^0$ mesons 
from their vacuum values in isospin symmetric matter and about 
31 MeV and 16 MeV for $D^+$ and $D^0$ mesons in the isospin 
asymmetric matter with $\eta$=0.5.
The mass drop of $D^+$ meson of 23 MeV in symmetric nuclear matter
at zero temperature may be compared with the value of about 43 MeV,
which is the drop in $D^+$ mass in the Born approximation of the 
calculation of the coupled channel approach of Ref. \cite{mizutani6}, 
when only the Weinberg Tomozawa interaction is included. 
In the presence of the scalar exchange term as well, the
drop in the $D^+$ mass in the present investigation is seen to be
about 144 MeV, which may be compared to the drop in the mass
of the $D^+$ meson with Weinberg-Tomozawa as well as scalar 
interaction of about 60 MeV in Ref. \cite{mizutani6}.
In the present investigation, we observe the scalar interaction
to be much more dominant than in Ref. \cite{mizutani6}.
The contributions to the energy of $D^+$ and $D^0$ mesons due to 
the explicit symmetry breaking term (scalar meson 
exchange term) are observed to increase with temperature 
upto a particular value of temperature, above which it is seen to 
decrease with temperature. This is a reflection of the 
behaviour of the scalar fields $\sigma$ and $\zeta$ shown in
figures \ref{fig1} and \ref{fig2}, whose magnitudes are seen 
to increase with temperature upto a particular value, above which
there is seen to be a decrease in their magnitudes. 
The range term is seen to give an increase in the $D^+$ and
$D^0$ masses and at $\rho_B=\rho_0$, there is seen to be
a rise of about 71 MeV at zero temperature for $\eta$=0.
This value is modified to about 67 MeV at T=150 MeV and
about 70 MeV at T=175 MeV. For a density of $\rho_B=4\rho_0$
and at zero temperature, the
drop in the D-meson masses in the symmetric nuclear matter
is seen to be about 27 MeV due to the range term. For this density, 
the contribution
of the $d_1$ and $d_2$ terms dominate over the first range term
thus leading to an overall drop of the D-meson masses in the
medium of about 347 MeV in the symmetric nuclear matter
at zero temperature. 

It may be noted that the difference in the in-medium energy of 
$D^{+}$ and $D^{-}$ mesons arises from Weinberg-Tomozawa term only 
(see  equations (\ref{selfd}) and (\ref{selfdbar})) whereas the 
the scalar meson exchange term and the range terms have 
identical contributions to the energy of $D^{+}$ and $D^{-}$ mesons. 
The repulsive contribution for the $D^-$ and $\bar {D^0}$ mesons
for the Weinberg-Tomozawa term as compared to the attractive 
contribution for the $D^+$ and $D^0$ mesons, gives a rise
in the masses of the $\bar D$ mesons (plotted in
figure \ref{fig6}) as compared to the masses of the $D$ mesons
shown in figure \ref{fig5}. At zero density, the contribution
from the Weinberg-Tomozawa term is zero and hence the
modifications for the $D^+$ and $D^-$ are identical,
both having negligible changes in their masses due to the
cancelling effects of the attractive scalar term and the 
repulsive range terms. At nuclear matter saturation density,
when one takes into account only the contribution from the 
Weinberg-Tomozawa term, there is seen to be an increase 
in the mass of the $D^-$ meson of about 24 MeV.
However, when we consider the contributions due
to all the individual terms, the attractive scalar interaction 
is seen to dominate over the
repulsive Weinberg-Tomozawa and the range terms, therby 
giving a drop of the $D^-$ mass of about 27 MeV in the
symmetric nuclear matter for $\rho_B=\rho_0$. 
However, for higher densities, the $d_1$ and $d_2$ terms
of the range term, which are both attractive, give
a further drop of the mass of the $D^-$ meson
and for density, $\rho_B=4\rho_0$, the overall drop
is seen to be about 162 MeV for $\eta$=0.


\begin{figure}
\includegraphics[width=16cm,height=16cm]{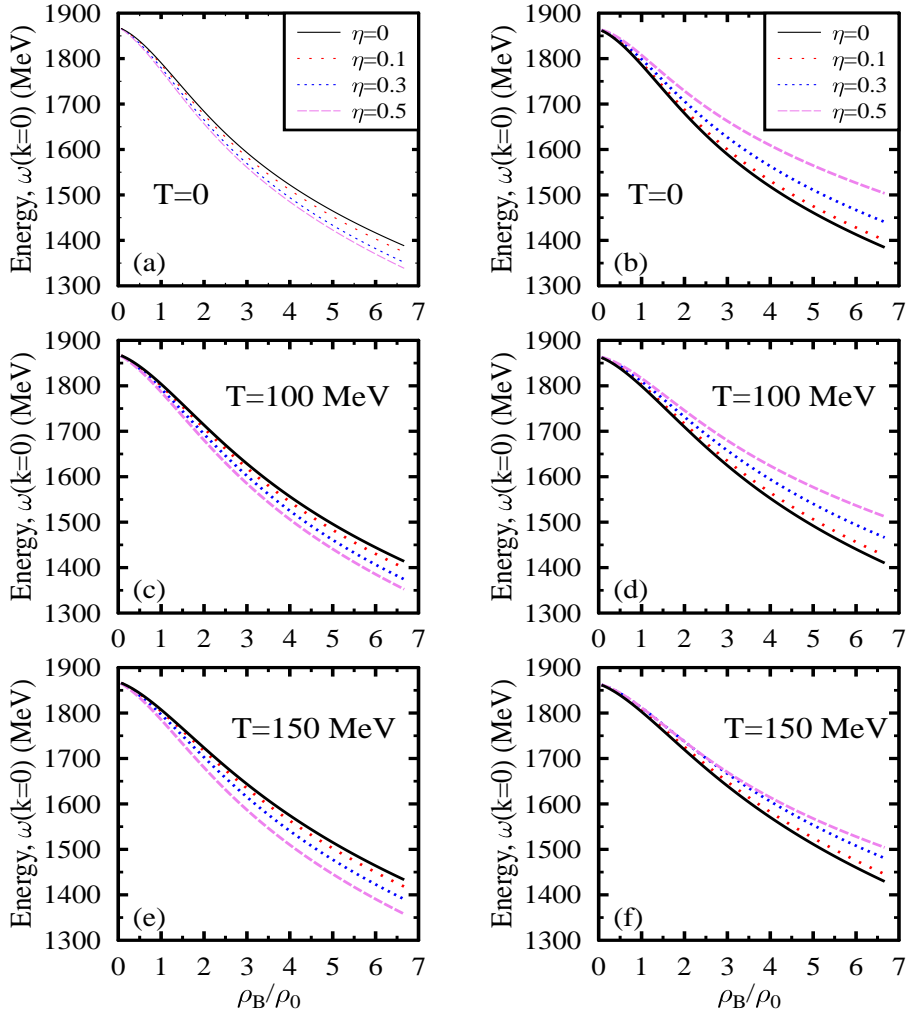} 
\caption{(Color online) The energies of $D^{+}$ meson ((a),(c) and (e)) 
and of $D^{0}$ meson ((b),(d) and (f)), at momentum $k = 0$, versus 
the baryon density (in units of nuclear saturation density), 
$\rho_{B}/\rho_{0}$, for different values of the isospin asymmetry 
parameter ($\eta = 0, 0.1, 0.3, 0.5$) and for given values of temperature 
(T = 0, 100 MeV and 150 MeV). The values of parameters $d_{1}$ and $d_{2}$ are calculated from $KN$ scattering lengths
in I = 0 and I = 1 channels.} 
\label{fig7}
\end{figure}

We study the density dependence of $D$ and $\bar{D}$ masses at 
finite temperatures at selected values of the isospin asymmetric paramater, 
$\eta$ and compare the results with the zero temperature case \cite{amarind}. 
The isospin symmetric part of the Weinberg-Tomazawa term gives a drop
of the $D$ mass, as can be
seen from the expressions of the self-energies of the $D$ and $\bar D$ 
mesons given by equations (\ref{selfd}) and (\ref{selfdbar}). 
However, the isospin asymmetric part of this term is seen to give 
a mass splitting for the $D^+$ and $D^0$, given by the second term
of the Weinberg-Tomazawa term, giving a further drop of the $D^+$ mass,
whereas the asymmetry reduces the drop of the mass of $D^0$.
For the $\bar D$ mesons, the isospin symmetric part of the Weinberg  
Tomazawa term gives an increase in the mass and the isospin asymmetric 
contribution of this term gives a further rise in the mass of the
$D^-$ mass, whereas it reduces the increase of the $\bar {D^0}$ 
mass in the asymmetric nuclear medium. For both $D$ and $\bar D$ mesons 
in the symmetric nuclear matter, the scalar meson interaction is 
attractive and identical (except for a small difference due to 
the difference in the vacuum masses of $D^\pm$ and $D^0 
(\bar {D^0})$ mesons). One might observe from the expressions 
of the $D (\bar D)$ self energies, given by equations (\ref{selfd})
and (\ref{selfdbar}) that the non-zero value of $\delta$ meson
arising due to the isospin asymmtery in the medium gives a drop
in the masses of $D^+ (D^-)$ in the asymmteric nuclear matter,
whereas this interaction is repulsive for $D^0 (\bar {D^0})$ mesons. 
The contributions to the $D$ and $\bar D$ masses due to the range
terms are given by the last three terms of the self energies
given by equations (\ref{selfd}) and (\ref{selfdbar}).
The first of these range terms is repulsive, whereas the second
and third terms are attractive, when isospin asymmtery is not
taken into account. However, due to a nonzero value of the 
$\delta$ field arising from isospin asymmtery in the medium,
the $\delta$ term of the first range term leads an increase 
in the masses of the $D^+(D^-)$ and a drop in the masses of $D^0 
(\bar {D^0})$ mesons.  
The second of the range terms (the $d_1$ term) is
attractive and gives identical mass drops for $D^+$ and $D^0$ in the
$D$ doublet as well as for $D^-$ and $\bar D^0$ in the $\bar D$ doublet.
This term is proportional to $(\rho_s^p+\rho_s^n)$, which turns
out to be different for the isospin asymmetric case
as compared to the isospin symmetric nuclear matter,
due to the presence of the $\delta$ meson.
This is because the equations of motion for the scalar fields
for the two situations (with/without $\delta$ mesons) give different values
for the mean field, $\sigma$ ($\sim ({\rho_s}^p +{\rho_s}^n)$).
The last term of the range term (the $d_2$ term) has a negative
contribution for the energies of $D^+$ and $D^0$ mesons
as well as for $D^-$ and $\bar {D^0}$ mesons for the isospin
symmetric matter. The isospin asymmetric part arising from the
($({\rho_s}^n -{\rho_s} ^p)$) term of the $d_2$ term has a further
drop in the masses for $D^\pm$ mesons,
whereas it increases the masses of the $D^0$ and $\bar{D^0}$ mesons
from their isospin symmetric values.
In Fig. \ref{fig7}, we show the variation of the energy of the $D$ mesons 
($D^{+}, D^{0}$) at zero momentum with baryon density $\rho_{B}$ for
different values of isospin asymmetry parameter $\eta$ and with the values 
of temperature as T = 0, 100, 150 MeV. The isospin-asymmetry in the 
medium is seen to give an increase in the  $D^{0}$ mass and a drop in 
the $D^{+}$ mass as compared to the isospin symmetric ($\eta = 0$) case. 
This is observed both for zero temperature \cite{amarind} and finite 
temperature cases. 
At nuclear matter saturation density, $\rho_{B} = 
\rho_{0}$, the drop  in the mass of $D^{+}$ meson from its vacuum value 
(1869 MeV) is 78 MeV for zero temperature case in isospin symmetric medium. 
At a density of $4\rho_{0}$, this drop in the mass of $D^{+}$ meson is 
seen to be about 347 MeV for $\eta$=0. At finite temperatures, 
the drop in the mass of $D^{+}$ meson for a given value of isospin asymmetry
decreases as compared to the zero temperature case. For example, at 
nuclear saturation density, $\rho_{0}$, the drop 
in the mass of $D^{+}$ meson turns out to be 72, 65 and 62 MeV at a 
temperature of T = 50, 100 and 150 MeV respectively for the isospin symmetric
matter. At a higher density of $\rho_B=4\rho_0$, zero temperature value for
the $D^+$ mass drop of about 347 MeV is modified to 336, 313 and 294 MeV
for T = 50, 100 and 150 MeV respectively. 
Thus the masses of $D$ mesons 
at finite temperatures and finite densities are observed to be larger than
the values at zero temperature case. This is because of the increase in 
the magnitudes of the scalar fields $\sigma$ and $\zeta$ with temperature 
at finite densities as mentioned earlier. The same behaviour remains for 
the isospin asymmetric matter. This behaviour of the nucleons and hence 
of the D-mesons with temperature was also observed earlier for symmetric 
nuclear matter at finite temperatures within the chiral effective model 
\cite{amdmeson}. The drop in the mass of $D^{+}$ meson is seen to be larger 
as we increase the value of the isospin asymmetry parameter. We observe that 
as we change $\eta$ from 0 to 0.5, then the drop in the mass of $D^{+}$ 
meson is 95 MeV and 384 MeV at densities of $\rho_{0}$ and $4\rho_{0}$
respectively, for the zero temperature case. At $ T = 50 $ MeV these 
values change to 89 MeV at $\rho_{0}$ and 377 MeV at a density of 
$4\rho_{0}$. For a baryon density, $\rho_B=\rho_0 (4\rho_0)$,
the drop in the $D^+$ mass is 83 (363) MeV at $T = 100$ MeV, 
and 84 (359) MeV at T=150 MeV. We thus observe that for a baryon density
of $\rho_0$, for $\eta$=0.5, the drop in the $D^+$ mass from its value 
for $\eta$=0 is 17, 18 and 22 MeV for T=0, 100 and 150 MeV respectively,
and for $\rho_B=4\rho_0$, these values are modifed to 37, 50 and 65 MeV
respectively. Thus we observe that for a given value of density, as we move 
from $\eta = 0$ to $\eta = 0.5$ the drop in mass of $D^{+}$ mesons is 
larger at higher temperatures.
\begin{figure}
\includegraphics[width=16cm,height=16cm]{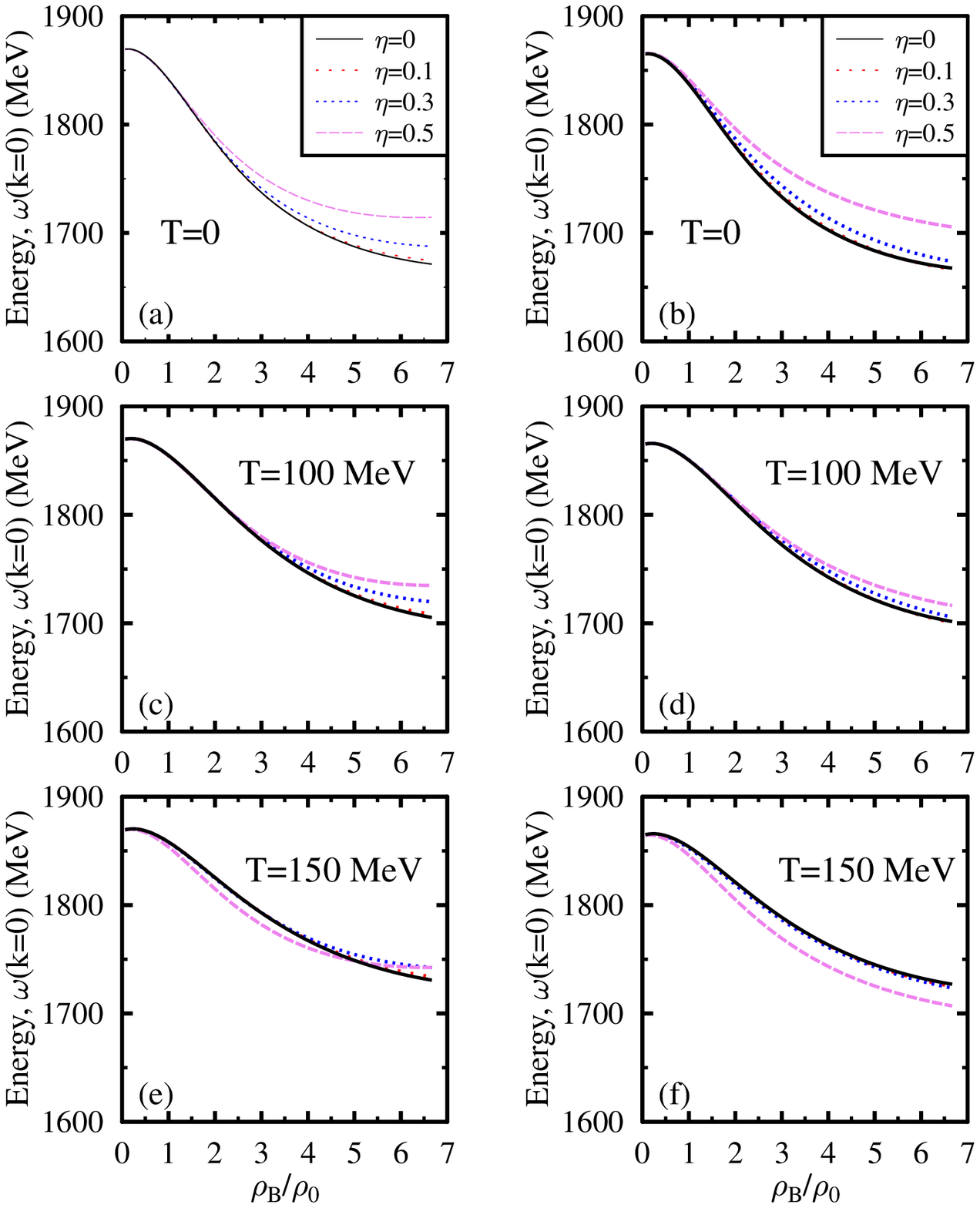} 
\caption{(Color online) The energies of $D^{-}$ meson ((a),(c) and (e)) 
and of $\bar {D^0}$ meson ((b),(d) and (f)), at momentum $k = 0$, versus 
the baryon density, expressed in units of nuclear matter saturation density, 
$\rho_{B}/\rho_{0}$, for different values of the isospin asymmetry 
parameter ($\eta = 0, 0.1, 0.3, 0.5$) and for a given temperature 
(T = 0, 100 MeV and 150 MeV). The parameters $d_{1}$ and $d_{2}$
are determined from $KN$ scattering lengths in I = 0 and I = 1 channels.} 
\label{fig8}
\end{figure}

The mass of the $D^0$ meson drops with density as can be seen from figure 
\ref{fig7}. The drop in the mass of $D^{0}$ meson at density $\rho_{0}$ 
from its vacuum value (1864.5 MeV) is 78, 65, 62 MeV for temperature 
$T = 0, 100, 150$ MeV 
respectively at $\eta = 0$. At density $4\rho_{0}$ and isospin-asymmetry 
parameter $\eta = 0$, these values become 347, 313 and  
294 MeV for temperatures T=0,100 and 150 MeV respectively. 
For the $D^{0}$ meson, there is seen to be an increase in the mass 
as we move from isospin symmetric to isospin asymmetric medium. 
For example, at zero temperature and baryon density equal to $\rho_{0}$ 
and $4\rho_{0}$, the rise in the masses of the $D^{0}$ meson are 21 and 91 
MeV respectively as we move from isospin-symmetric medium ($\eta = 0$) 
to isospin-asymmetric medium ($\eta = 0.5$). At a temperature of 100 MeV, 
these values become 17 MeV at $\rho_{0}$ and 72 MeV at $4\rho_{0}$. 
For T=150 MeV, these values become 9 MeV at $\rho_{0}$ and 
44 MeV  at $4\rho_{0}$. Thus for $D^{0}$ mesons, the rise in the mass, 
is seen to be lowered at higher temperatures as we move from $\eta$ = 0 to 
$\eta$ = 0.5. The strong isospin dependence of the $D^+$ and $D^0$ meson 
masses should show up in observables such as their production as well as 
flow in asymmetric heavy-ion collisions planned at the future facility 
at FAIR, GSI.

Fig.\ref{fig8} shows the results for the density dependence of the 
energies of the $\bar{D}$ mesons at zero momentum at values of the
temperature, $ T = 0, 100, 150$ MeV. There is seen to be a drop 
of the masses of both the $D^-$ and $\bar {D^0}$ with density.
This is due to the dominance of the attractive scalar exchange 
contribution as well as the range terms (which becomes attractive 
above a density of about 2--2.5 times the nuclear matter saturation 
density) over the repulsive Weinberg-Tomozawa interaction \cite{amarind}. 
It is observed that the drop in the mass of $D^{-}$ and $\bar{D}^{0}$ 
in isospin symmetric nuclear matter is 27.2 MeV at $\rho_{0}$ 
and 162 MeV at $4\rho_{0}$, at zero temperature, from their vacuum 
values. As we go to higher temperatures, the drop in the masses of
$D^{-}$ and $\bar{D}^{0}$ mesons decreases. For example, at $\eta$ = 0 
and $\rho_{B} = \rho_{0}$ the drop in the mass of $D^{-}$ meson is 
20.9, 14.3 and 10.8 MeV at a temperature of 50, 100 and 150 MeV respectively. 
For temperature, T=0,50,100 and 150 MeV, at $\rho_B=\rho_0$,
the values of the drop in the $D^-$ mass are seen to be 
27.2, 21.4, 14.6 and 15.5 MeV respectively,
and the drop in the $\bar {D^0}$ mass are modifed
from about 27, 21, 14.2 and 11 MeV to 23, 19, 14 and
19 MeV, when $\eta$ is changed from 0 to 0.5.
The masses of $D^{-}$ and $\bar{D}^{0}$ mesons are observed to have 
negligible dependence on the isospin asymmetry upto a density of 
$\rho_{B}=\rho_{0}$. However, at high densities there is seen to be 
appreciable dependence of these masses on the parameter, $\eta$. 
As we change $\eta$ from 0 to 0.5, at $\rho_B=4\rho_0$, the drop 
in the mass of the $D^-$ meson is modified from 
162 MeV to 139 MeV at zero temperature.
At higher temperatures, T= 50, 100 and 150 MeV and at the
density of $4\rho_0$, when we change $\eta$ from 0 to 0.5,
the values of the $D^-$ mass drop are modifed from 149, 123 and 
102 MeV to 128, 111 and 107 MeV respectively.
It is seen that, at high densities there is an increase 
in the masses of both $D^{-}$ and $\bar{D}^{0}$ mesons in isospin 
asymmetric medium as compared to those in the isospin symmetric 
nuclear matter for temperatures T=0, 50 and 100 MeV. However, 
at $T = 150$ MeV, it is observed that for densities upto about 
4.5$\rho_0$, the mass of $D^{-}$ meson is higher in the isospin 
symmetric matter as compared to in the isospin asymmetric matter 
with $\eta$ = 0.5. It is also seen that the modifications in the 
masses of $D^{-}$ mesons is negligible as we change $\eta$ from
0 to 0.3 upto a density of about 4$\rho_0$. 
For the $\bar{D}^{0}$ meson, one sees that the isospin dependence 
is negligible upto $\eta$=0.3. This is because the drop in the mass 
of $\bar{D}^{0}$ mesons due to isospin asymmetry given by 
Weinberg-Tomozawa term almost cancels with the increase due to 
the scalar and range terms as we go from $ \eta = 0$ to 
$ \eta = 0.3$. At zero temperature \cite{amarind} as well as for 
temperatures T=50 and 100 MeV, there is seen to be an increase 
in the mass of the 
$\bar{D}$ mesons ($D^-$,$\bar {D^0}$) as we go from isospin symmetric 
medium to the isospin asymmetric medium. This is because 
for T=0,50 and 100 MeV, the increase in mass of 
$\bar D$  given by the scalar exchange and the range terms 
dominate over the drop given by the Weinberg Tomozawa term as 
we go from symmetric nuclear medium ($\eta = 0$) to isospin asymmetric 
nuclear medium ($\eta$ =0.1,0.3 0.5). However, at $T = 150$ MeV,
for $\eta$ =0.5, the drop given by Weinberg term dominates over 
the rise given by scalar and range terms for $\bar {D^0}$ and upto
a density of about 4.5$\rho_0$ for $D^-$, and therefore mass of 
$\bar{D}$ meson decreases as we go from symmetric nuclear medium 
to isospin asymmetric nuclear medium in these density regimes.

The medium modifications of the masses of $D$ and $\bar{D}$ mesons 
in the present
investigation are due to the interactions with the nucleons and scalar 
mesons $\sigma$, $\zeta$ and $\delta$ in the hot nuclear medium. The 
values of the scalar fields in the medium are obtained by solving the 
the equations of motion for the scalar fields and the dilaton field 
$\chi$, given by the coupled equations (9) to (12).
The temperature and density dependence of the dilaton field
$\chi$ is seen to be negligible and thus the changes of the 
values of the scalar fields are observed to be marginal for the 
present investigation when 
the medium dependence of the $\chi$ field is taken into account
as compared to when its medium dependence is not taken into account 
(the so-called frozen glueball approximation where the value of
$\chi$ is taken to be its vacuum value). This leads to the 
modifications of the $D$ and $\bar D$ meson masses as marginal 
as compared to the case when the medium dependence 
of the dilaton field is not taken into account. For example,  
for temperature T = 0, the drop in the mass of $D^{+}$ mesons 
in isospin symmetric nuclear medium is about $81$ MeV and $364$ MeV 
at $\rho_{B} = \rho_{0}$ and $4\rho_{0}$ respectively in the frozen
glueball approximation, which may be compared to the values of 
mass drop as $78$ MeV and $347$ MeV in the present investigation, 
when we take into account the effect of variation of the dilaton field 
with density. Hence one observes the difference between them to be
marginal, of the order of about 5\%. Similarly, the mass drop in the 
$D^{-}$ meson for the isospin asymmetric matter at T=0 in the frozen glueball
approximation is observed to be about $30$ MeV and $184$ MeV at 
$\rho_{B} = \rho_{0}$ and $4\rho_{0}$ respectively. These are
different from the values 27 MeV and 162 MeV of the present calculations 
with medium dependent dilaton field, by about 10\%. 

In the present calculations, we have used the value of 
decay constant, $f_{D} = 135$ MeV. 
In isospin symmetry nuclear medium, for temperature T =0, we observe that 
the values of mass drops for $D^{+}$ and $D^{-}$ mesons are about
42 MeV and 4 MeV respectively at nuclear saturation density 
when we set $f_D$=157 MeV \cite{fd157}. These values
may be compared to the values of 78 MeV and 27 MeV when the
$f_D$ is taken to be 135 MeV. Hence by modifying the value of $f_D$
by about 15$\%$ leads to modifications of the mass shifts of $D^+$ 
and $D^-$ by about 46$\%$ and 85$\%$ at density, $\rho_0$ for symmetric 
nuclear matter at zero temperature. The mass drops
for the value of the density as 4$\rho_0$ for symmetric nuclear matter 
at T=0, are modified from 347 MeV and 162 MeV to the values 238 MeV 
and 93 MeV for the $D^+$ and $D^-$ respectively, when we change the
value of $f_D$ from 135 MeV to 157 MeV. Hence, for $\rho_B=4\rho_0$,
the modifications for the $D^+(D^-)$ masses are about 30$\%$ and 40$\%$
when we change the value of the D-meson decay constant. Hence, there
seems to be appreciable dependence of the mass shifts of the 
$D(\bar D)$ mesons with the D-meson decay constant.

\begin{figure}
\includegraphics[width=16cm,height=16cm]{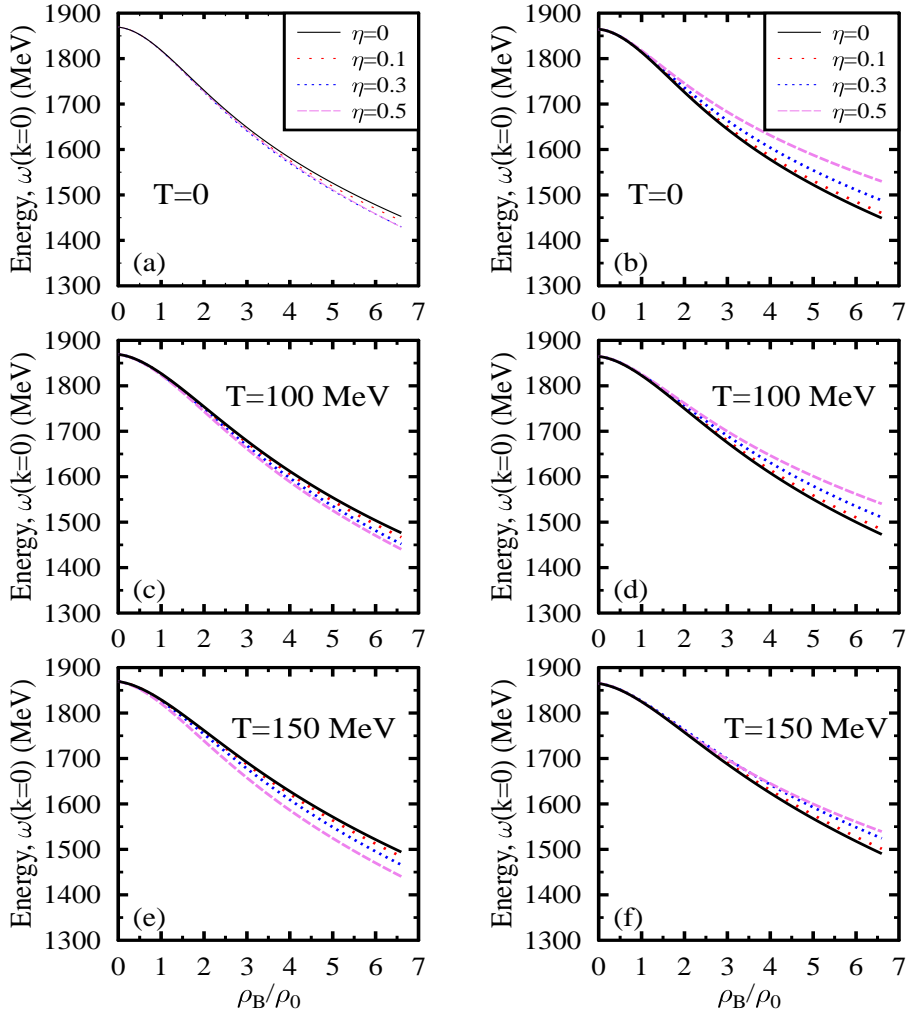} 
\caption{(Color online) The energies of $D^{+}$ meson ((a),(c) and (e)) 
and of $D^{0}$ meson ((b),(d) and (f)), at momentum $k = 0$, versus 
the baryon density (in units of nuclear saturation density), 
$\rho_{B}/\rho_{0}$, for different values of the isospin asymmetry 
parameter ($\eta = 0, 0.1, 0.3, 0.5$) and for given values of temperature 
(T = 0, 100 MeV and 150 MeV). The values of parameters $d_{1}$ and $d_{2}$ are calculated from $DN$ scattering lengths
in I = 0 and I = 1 channels.} 
\label{fig9}
\end{figure}
\begin{figure}
\includegraphics[width=16cm,height=16cm]{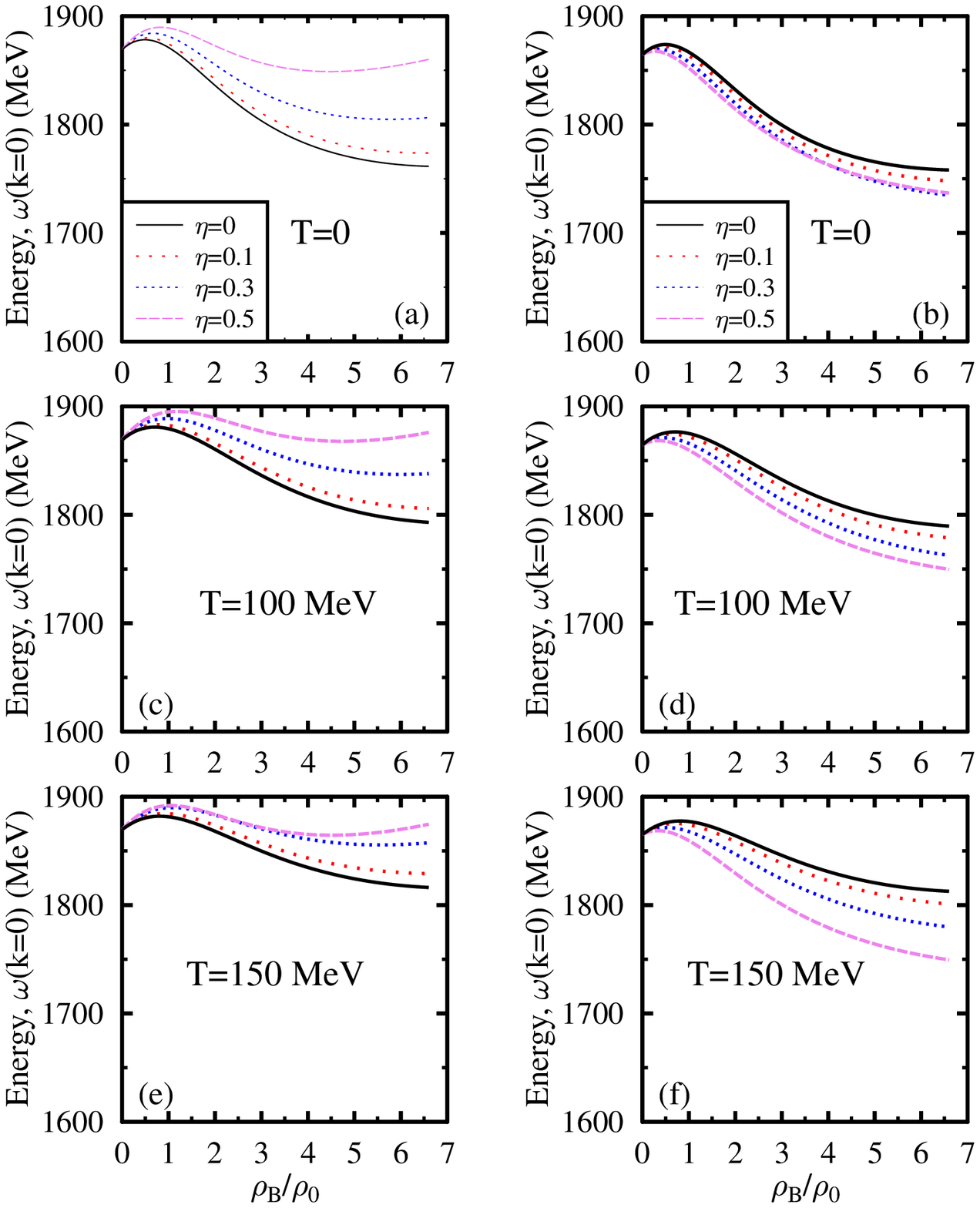} 
\caption{(Color online)  The energies of $D^{-}$ meson ((a),(c) and (e)) 
and of $\bar {D^0}$ meson ((b),(d) and (f)), at momentum $k = 0$, versus 
the baryon density, expressed in units of nuclear saturation density, 
$\rho_{B}/\rho_{0}$, for different values of the isospin asymmetry 
parameter ($\eta = 0, 0.1, 0.3, 0.5$) and for given values of temperature 
(T = 0, 100 MeV and 150 MeV). The values of parameters $d_{1}$ and 
$d_{2}$ are calculated from $DN$ scattering lengths in I = 0 and 
I = 1 channels.} 
\label{fig10}
\end{figure}

We next examine how the masses of the $D$ and $\bar{D}$ mesons change 
if we determine the parameters $d_{1}$ and $d_{2}$ from DN scattering 
lengths calculated to be $-0.43$ fm and $-0.41$ fm in the I = 0 and 
I = 1 channels respectively, in a coupled channel approach  \cite{MK}.
In figures \ref{fig9} and \ref{fig10}, we show the variation of 
the energies of $D$ and $\bar D$ mesons respectively, at zero momentum, 
with baryon density $\rho_{B}$ for different values of the isospin 
asymmetry parameter $\eta$ and for the values of the temperature 
as T = 0, 100, 150 MeV. The values of $d_{1}$ and 
$d_{2}$ parameters determined from these values of the
DN scattering lengths turn out to be $8.95/m_{D}$ and 
$0.52/m_{D}$ respectively, which can be expressed in terms of the
mass of the kaon as $2.385/m_K$ and $0.14/m_K$. 
These values of parameters are smaller than the values 
of $d_{1}$ and $d_{2}$ as $2.56/m_{K}$ and $0.73/m_{K}$ 
respectively, when determined from the KN scattering lengths 
in $I =0$ and $I = 1$ channels. Since both of these terms  
are attractive, the masses of the $D$ ($\bar D$) 
mesons turn out to have a smaller drop when these parameters 
are fitted from DN scattering lengths as compared to when these 
are fitted from the KN scattering lengths.

With the set of values of $d_{1}$ and $d_{2}$ parameters as fitted from
the DN scattering lengths, the drop in the masses of $D^{+}$ meson 
in the isospin symmetric nuclear medium at nuclear matter saturation 
density $\rho_{0}$ turns out to be $51, 42$ and $40$ MeV, at the values 
of temperature as, $T = 0, 100$ and $150$ MeV respectively. These may
be compared to the values of the mass drop of $D^+$ meson
of 78, 65 and 62 MeV for T=0, 100 and 150 MeV, respectively,
when the parameters are fitted from the KN scattering lengths.
For $D^{0}$ mesons the values of mass drop are $50, 42$ and 
$40$ MeV at T=0, 100 and 150 MeV respectively, when $d_1$ and
$d_2$ are fitted from the DN scattering lengths. These may be
compared to the values of mass drop of $D^0$ meson of about
78, 65 and 62 MeV, when these are fitted to the KN scattering lengths.  
At higher densities, the $d_1$ and $d_2$ terms become more dominant
and overcome the repulsive interaction of the first range term,
leading to a drop of the D-meson masses due to the range term
as well. At a density of $\rho_B=4\rho_0$, the drop in the $D^+$ mass
is observed to be about 286 MeV, 256 MeV, 240 MeV for T=0, 100 MeV
and 150 MeV, which may be compared to the values of 347 MeV, 313 MeV
and 294 MeV for T=0,100, 150 MeV, when the parameters $d_1$ and $d_2$ 
are fitted from the KN scattering lengths.  

We plot the masses of the $\bar{D}$ mesons in figure \ref{fig10},
with the values of $d_{1}$ and $d_{2}$ are fitted from $DN$ scattering 
lengths \cite{MK}. At the nuclear saturation density $\rho_{0}$, 
the value of mass of $D^{-} (\bar {D^0})$ meson in isospin symmetric 
nuclear medium, is observed to increase by 2, 10 (10.5) and 12 (12.5) MeV 
at $T = 0, 100$ and $150$ MeV, respectively. However, at
higher densities, the $d_1$ and $d_2$ terms become more
dominant thus leading to a drop of the $D^- (\bar {D^0})$ masses
in the nuclear matter. For $\rho_B=4\rho_0$, 
the mass of $D^- (\bar {D^0})$ meson is seen to decrease
by 87 (86) MeV, 52 (51.5) MeV and 34 (33.6) MeV respectively.
These may be compared to the results of the in-medium masses
of $D^- (\bar {D^0})$ meson, when $d_1$ and $d_2$ are fitted from the
KN scattering lengths. In the latter case, as already mentioned
for T=0, 100 MeV and 150 MeV and for $\rho_B=4\rho_0$, there is 
seen to be a drop of $D^- (\bar {D^0})$ mass of 27.2 (27.2)
14.3 (14.2) and 
10.8 (10.8) MeV at $\rho_B=\rho_0$  and 162 (162)
122.6 (122) and 101.7 (101.2) MeV at $\rho_B=4\rho_0$.

As mentioned earlier, the in-medium mass of $D^{+}$ meson decreases 
with increase in the isospin asymmetry of the nuclear medium. However, 
with values of $d_{1}$ and $d_{2}$ parameters fitted from DN scattering 
lengths, the decrease in the mass with isospin asymmetry of the 
nuclear medium is observed to be smaller than that when these
parameters are fitted from the KN scattering lengths. For example, 
at $\rho_{B} = 4\rho_{0}$ and $T = 0$, as we move from $\eta = 0$ 
to $\eta = 0.3$, the in-medium masses of $D^{+}$ mesons decrease 
by $27$ MeV with the values from the KN scattaring lengths 
and by $13$ MeV with the set of values of $d_{1}$ and $d_{2}$ parameters
fitted from the DN scattering lengths.
For the $D^{+}$ meson, the $d_{1}$ term gives a rise in the mass 
whereas the $d_{2}$ term gives drop in the mass when one has 
a nonzero value of the asymmetry parameter as compared to the
symmetric nuclear matter.  Also, the first range term  gives an
increase due to isospin asymmetry, whereas both the Weinberg 
Tomozawa term and scalar meson exchange term give drop in the 
mass of the $D^{+}$ meson. Due to the drop arising from the $d_{2}$ term, 
Weinberg Tomozawa term and the scalar term dominating over the
the increase due to the $d_1$ term, the mass of $D^{+}$ meson 
decreases in the isospin asymmetric nuclear medium as compared
to the mass in symmetric matter.  
When we use the values of the parameters $d_{1}$ and $d_{2}$ as calculated
from the DN scattering lengths, then there is a smaller 
contribution to the drop of the mass due to $d_{2}$ term, as compared
to the drop due to isospin asymmetry arising to this term 
when the parameters $d_1$ and $d_2$ are calculated from the KN scattering 
lengths. This is due to the smaller value for the $d_2$ in the former
case. Therefore, with values of $d_{1}$ and $d_{2}$ parameters
calculated from the DN scattering lengths, the drop in 
the mass of $D^{+}$ meson with isospin asymmetry of the medium is
seen to be small. 

With the set of values of the parameters $d_{1}$ and $d_{2}$
as fitted from the KN scattering lengths, the mass of the $D^{-}$ 
meson is observed to increase with the isospin asymmetry 
of the medium. With the values of $d_{1}$ and $d_{2}$ parameters
as fitted from the DN scattering lengths, there is seen to be 
larger increase in the mass of the $D^{-}$ meson due to the isospin 
asymmetry of the nuclear medium. 
For example, at $\rho_{B} = 4\rho_{0}$ and $T = 0$, as we move from 
$\eta =0$ to $\eta = 0.3$, the mass of the $D^{-}$ meson is seen to
increase by about $7$ MeV when we use the values of $d_1$ and $d_2$ 
as fitted from the KN scattering lengths and by $32$ MeV with  
the values of $d_{1}$ and $d_{2}$ fitted from the DN scattering lengths.
For the $D^{-}$ meson, there is an increase in the mass due to
isospin asymmetry due to the Weinberg Tomozawa term, the first range term 
(term with coefficient ($-\frac{1}{f_{D}}$) and the $d_{1}$ term,
whereas the scalar term and $d_{2}$ term give drop in the mass 
of $D^{-}$ meson as compared to the symmetric nuclear matter. The net 
effect is that the mass of the $D^{-}$ meson increases with the isospin
asymmetry of the nuclear medium. With the values of $d_{1}$ and $d_{2}$
fitted from DN scattering lengths, because of the smaller value of $d_2$, 
the drop arising from the $d_{2}$ term due to isospin asymmetry 
in the medium is observed to be smaller than the case when the
parameters $d_1$ and $d_2$ are fitted from the KN scattering 
lengths (same as for $D^{+}$ mesons). There is seen to be larger
increase in the in-medium masses of $D^{-}$ mesons as a 
function of the isospin asymmetry of the nuclear medium
when the parameters are determined from the DN scattering 
lengths as compared to the KN scattering lengths. 

With the values of $d_{1}$ and $d_{2}$ calculated from KN scattering 
lengths, the mass of the $\bar{D^{0}}$ meson increases with the isospin 
asymmetry $\eta$ of the medium as shown in figure \ref{fig8}. However, 
if we use the values of $d_{1}$ and $d_{2}$, fitted from the $DN$ 
scattering lengths, then the mass of $\bar{D^{0}}$ mesons are seen to 
decrease with increase in the isospin asymmetry of the medium as 
shown in the figure \ref{fig10}. For example, at $\rho_{B} = 4\rho_{0}$ 
and $T = 0$, as we move from $\eta = 0$ to $\eta = 0.3$, the
$\bar{D^{0}}$ mass increases by about $10$ MeV when the parameters
$d_1$ and $d_2$ are fitted from the KN scattering length, whereas
the mass of $\bar {D^0}$ is seen to decrease by about $16$ MeV 
when $d_{1}$ and $d_{2}$ are calculated from the DN scattering
lengths. The reason is, as a function of isospin asymmetry of the medium 
the $d_{1}$ and $d_{2}$ terms give rise to the masses of 
$\bar{D^{0}}$ mesons and the first range term gives a drop in 
the $\bar {D^0}$ mass. The values of $d_1$ and $d_2$ are larger 
when fitted from the KN scattering lengths as compared to when
calculated from the DN scattering lengths. Hence in the former case, 
the the increase in $\bar {D^0}$ from isospin asymmetry arising from 
the $d_{1}$ and $d_{2}$ terms dominates over the drop given by first 
range term. Therefore, with isospin asymmetry of the medium the mass 
of $\bar{D^{0}}$ increases in the former situation. 
However, for the values of $d_{1}$ and $d_{2}$ parameters fitted 
from DN scattering lengths, the increase due to $d_1$ and $d_2$ terms
is dominated by first range term. The Weinberg term gives a drop and
the scalar term gives an increase in the mass of $\bar{D^{0}}$ meson 
with the isospin asymmetry of the nuclear medium. The net effect on
the mass of the $\bar{D^{0}}$ meson is a drop with 
isospin asymmetry, $\eta$, of the nuclear medium.  

The parameters $d_{1}$ and $d_{2}$ have the same effect on the masses 
of the $D^{0}$ and $\bar{D^{0}}$ mesons of giving an increase in their
mass in the isospin asymmetric medium as compared to the masses
in the symmetric nuclear matter whereas the first range term
gives a drop in their masses. 
The Weinberg Tomozawa term and the scalar meson exchange 
term lead to an increase in the mass of $D^{0}$ meson with isospin
asymmetry. As can be seen from figures \ref{fig7} and \ref{fig9},
the net effect on the $D^0$ meson mass is an increase
with isospin asymmetry, but the rise is less for the case when 
$d_1$ and $d_2$ are calculated from the DN scattering lengths
due to the smaller values of $d_1$ and $d_2$ as compared to
when these parameters are determined from the KN scattering lengths.
For example, at $\rho_{B} = 4\rho_{0}$ and $T = 0$, as we move 
from $\eta =0$ to $\eta = 0.3$, the in-medium mass of $D^{0}$ 
mesons increases by about $45$ MeV when the parameters are 
calculated from KN scattering lengths and by about $26$ MeV 
when fitted from the DN scattering lengths.

\begin{figure}
\includegraphics[width=16cm,height=16cm]{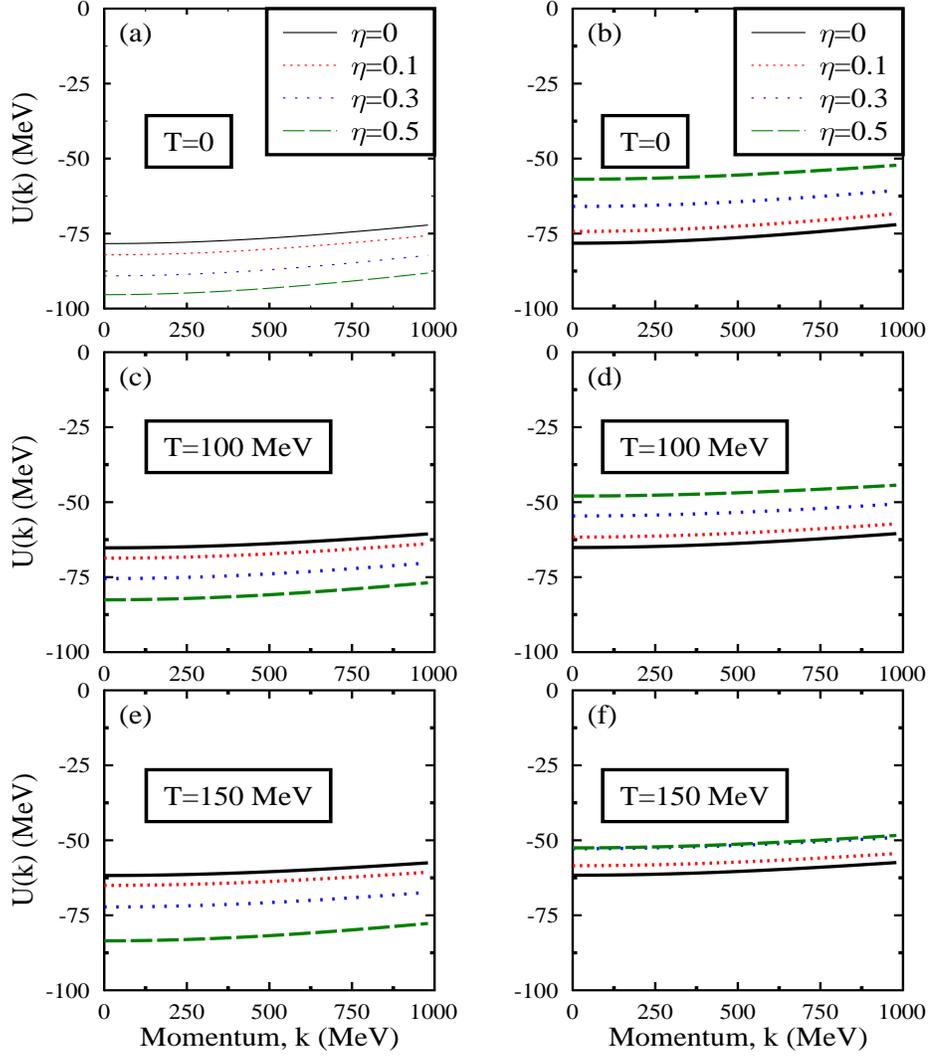} 
\caption{(Color online) The optical potential of $D^{+}$ meson 
((a),(c) and (e)) and of $D^{0}$ meson ((b),(d) and (f)), 
are plotted as functions of momentum for $\rho_{B}=\rho_0$, 
for different values of the isospin asymmetry 
parameter ($\eta = 0, 0.1, 0.3, 0.5$) and for given values 
of temperature (T = 0, 100 MeV and 150 MeV). The values of 
parameters $d_{1}$ and $d_{2}$ are calculated from $KN$ scattering lengths
in I = 0 and I = 1 channels.} 
\label{fig11}
\end{figure}
\begin{figure}
\includegraphics[width=16cm,height=16cm]{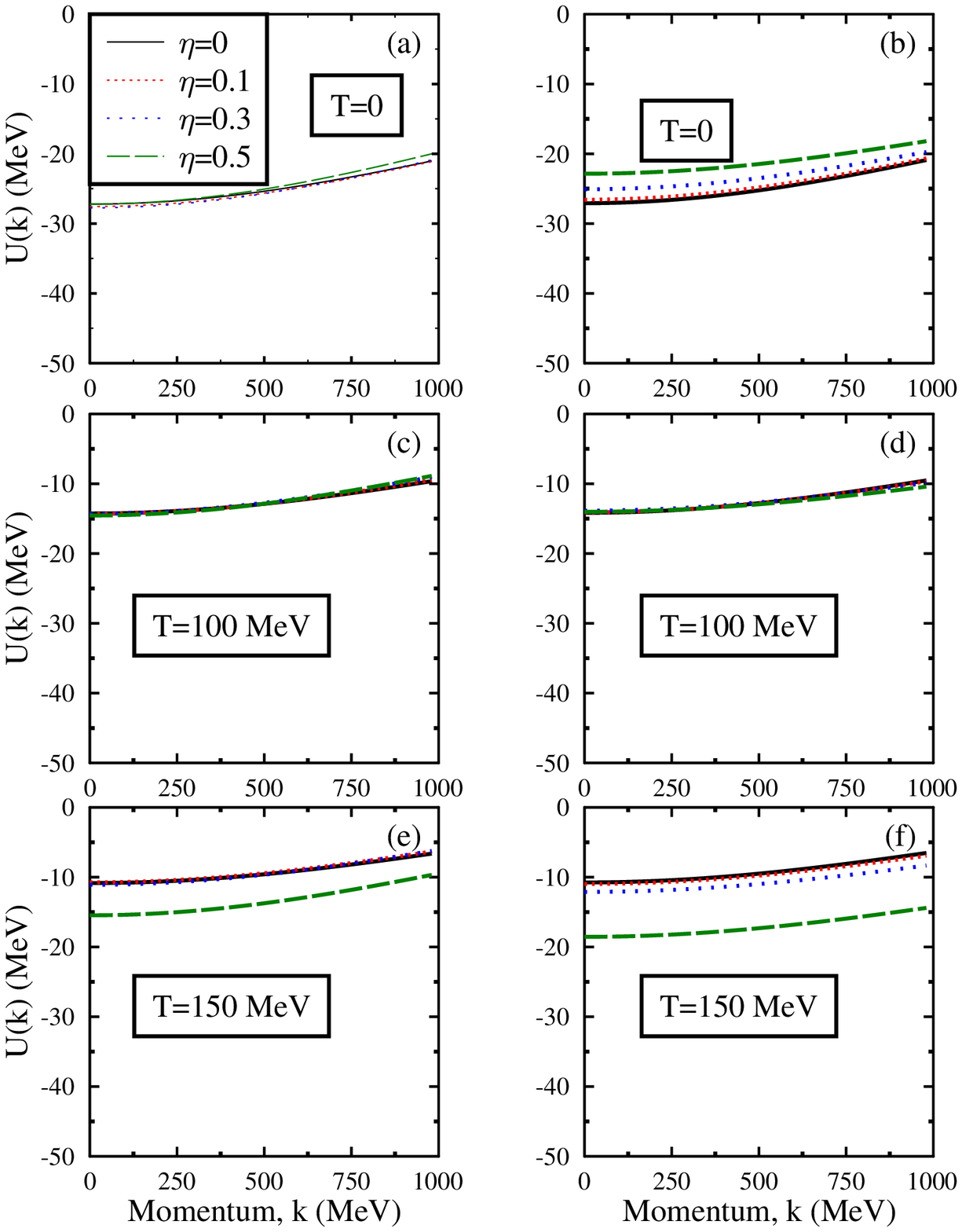} 
\caption{(Color online) The optical potential of $D^{-}$ meson 
((a),(c) and (e)) and of $\bar {D^{0}}$ meson ((b),(d) and (f)), 
are plotted as functions of momentum for $\rho_{B}=\rho_0$, 
for different values of the isospin asymmetry 
parameter ($\eta = 0, 0.1, 0.3, 0.5$) and for given values 
of temperature (T = 0, 100 MeV and 150 MeV). The values of parameters $d_{1}$ and $d_{2}$ are calculated from $KN$ scattering lengths
in I = 0 and I = 1 channels.} 
\label{fig12}
\end{figure}

\begin{figure}
\includegraphics[width=16cm,height=16cm]{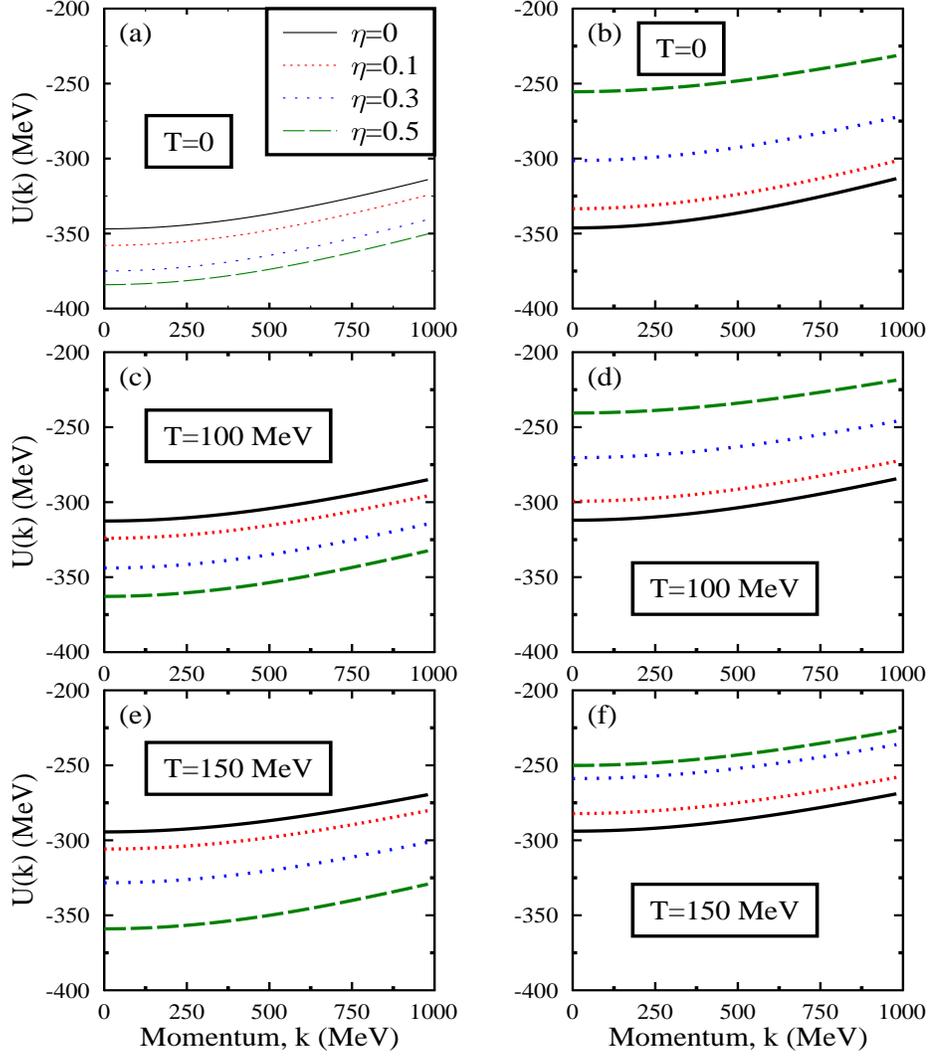} 
\caption{(Color online) The optical potential of $D^{+}$ meson 
((a),(c) and (e)) and of $D^{0}$ meson ((b),(d) and (f)), 
are plotted as functions of momentum for $\rho_{B}=4\rho_0$, 
for different values of the isospin asymmetry 
parameter ($\eta = 0, 0.1, 0.3, 0.5$) and for given values 
of temperature (T = 0, 100 MeV and 150 MeV). The values of parameters $d_{1}$ and $d_{2}$ are calculated from $KN$ scattering lengths
in I = 0 and I = 1 channels.} 
\label{fig13}
\end{figure}

\begin{figure}
\includegraphics[width=16cm,height=16cm]{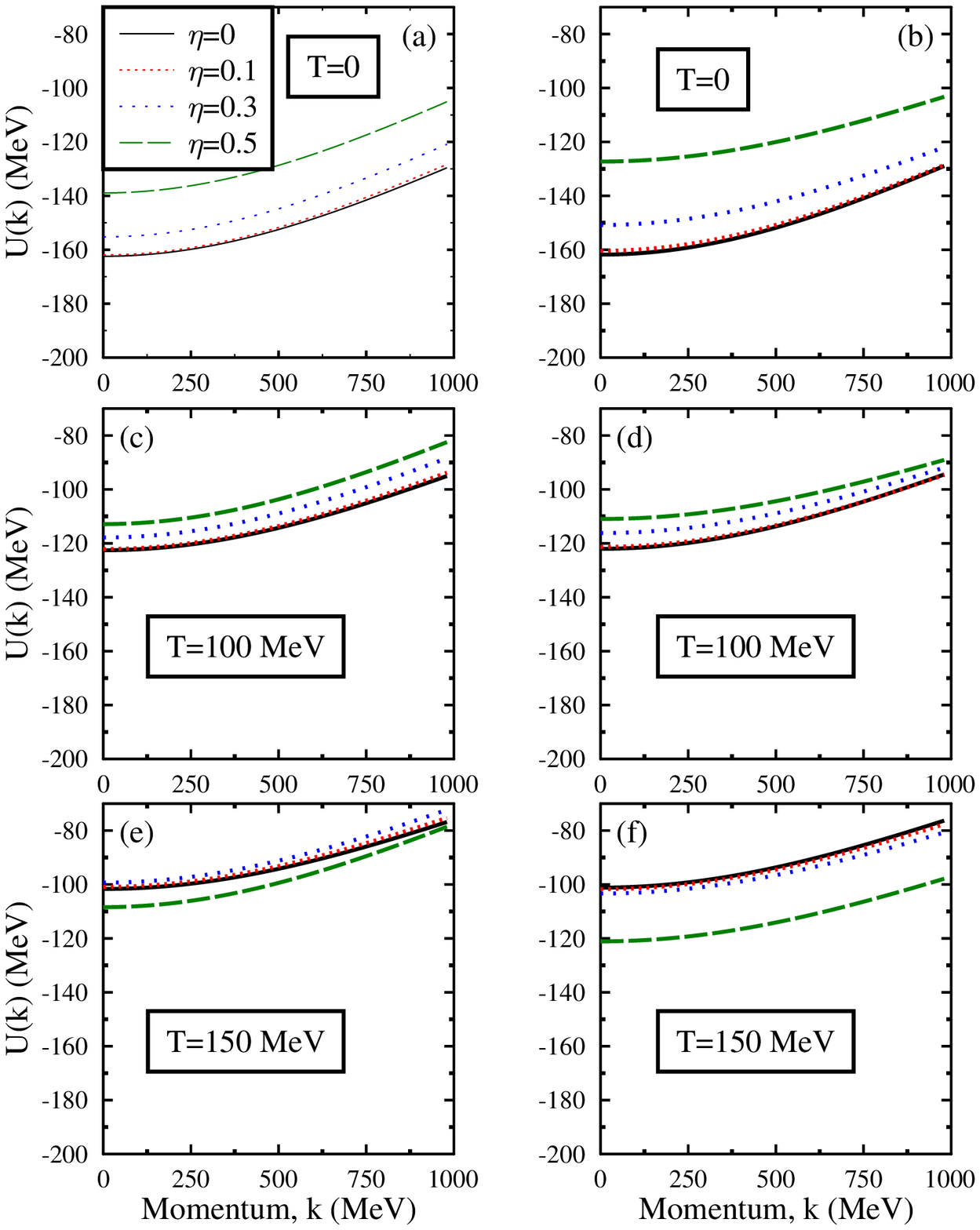} 
\caption{(Color online) The optical potential of $D^{-}$ meson 
((a),(c) and (e)) and of $\bar {D^{0}}$ meson ((b),(d) and (f)), 
are plotted as functions of momentum for $\rho_{B}=4\rho_0$, 
for different values of the isospin asymmetry 
parameter ($\eta = 0, 0.1, 0.3, 0.5$) and for given values 
of temperature (T = 0, 100 MeV and 150 MeV). The values of parameters 
$d_{1}$ and $d_{2}$ are calculated from $KN$ scattering lengths
in I = 0 and I = 1 channels.} 
\label{fig14}
\end{figure}

The mass modifications of $D$-mesons at finite density have been studied 
in the QCD sum rule  and the mass shift at nuclear matter saturation density 
was found to be about $-50$ MeV \cite{arata}. In the QMC model, the mass shift 
was around $-60$ MeV \cite{qmc}. In the present investigation, at finite 
densities the magnitude of the scalar fields first increases with increase 
in the temperature upto a particular value of temperature after which it 
starts decreasing. This behaviour is then reflected 
in the variation of the nucleon mass with temperature at finite densities. 
In QMC model, the behaviour of the scalar field is also seen to be the same 
as in the present model \cite{amqmc}. However, in the QMC model, the nucleon 
mass is seen to monotonically rise with temperature and there is no change 
in this trend observed even upto a temperature of about 250 MeV 
\cite{amqmc} unlike in the present chiral model or in the Walecka model 
\cite{frunstl1}. This is because $\sigma$ field in the QMC model 
is not as strong as in the chiral model or the Walecka model. In QMC model, 
there are contributions to the masses of nucleons from the thermal 
excitations of the quarks inside the nucleon bag. This contribution 
of quarks dominates over the $\sigma$ field in QMC model \cite{amqmc}.
The small attractive mass shift for the $\bar{D}$ mesons, obtained 
within our calculations are in favor of charmed mesic nuclei as 
suggested in the QMC model \cite{qmc}. This is, however, contrary to
a repulsive potential obtained for the $\bar D$ mesons in
the coupled channel approach \cite{mizutani8}.
In our investigation, if we do not take into consideration the effect 
of the range terms on the in-medium properties of $D$ mesons then at 
nuclear saturation density $\rho_{0}$ and temperature $T = 0$, 
the mass of $D^{+}$ and $D^{-}$ mesons drop by $144$ MeV and $96$ MeV 
respectively in isospin symmetric nuclear medium ($\eta = 0$). 
The large mass-shift shows the absence of repulsion due to total 
range term at nuclear saturation density.
However, in coupled channel approach of Ref.\cite{mizutani8}, 
a repulsive mass-shift of $11$ MeV is given for $\bar{D}$ mesons. 
Figures \ref{fig11} and \ref{fig13} show the isospin dependence
of the optical potentials for the $D$ mesons as functions of the momentum, 
for densities $\rho_{0}$ and $4\rho_{0}$ respectively and for values of
the temperature as $T = 0, 100, 150$ MeV.  These optical potentials
are plotted for the values of $d_1$ and $d_2$ calculated
from the KN scattering lengths. Figures \ref{fig12} 
and \ref{fig14} illustrate the optical potentials for the 
$\bar{D}$ doublet. The isospin dependence of optical potentials 
is seen to be quite significant for high densities for the D-meson doublet 
($D^{+}, D^{0}$) as compared to those for the $\bar D$ doublet.
This is a reflection of the strong isospin dependence of the
masses of the D-mesons as compared to the $\bar {\rm D}$ as has been
already illustrated in figures \ref{fig7} and \ref{fig8}.
For the $\bar{D}$ mesons, it is seen, from figure \ref{fig8}, 
that the masses of the the $D^{-}$ meson and $\bar{D}^{0}$ meson 
for a fixed value of the isospin asymmetry parameter, $\eta$ are 
very similar, an observation which was seen earlier for the
zero temperature case \cite{amarind}. These are reflected in
their optical potentials, plotted in figures \ref{fig12}
and \ref{fig14}, where one sees a maximum difference of 
about 5 MeV or so between $D^-$ and $\bar {D^0}$ for $\rho_B=\rho_0$
and about 10 -- 15 MeV for $\rho_B=4\rho_0$. 
The present investigations of the optical potentials for the
$D$ and $\bar D$ mesons show a much stronger dependence of
isospin asymmetry on the $D$ meson doublet, as compared to
that in the $\bar D$ meson doublet, as was already observed for
the zero temperature case. However, when the parameters $d_1$
and $d_2$ are fitted from the DN scattering lengths, then 
one sees a greater sensitivity to the isospin asymmetry
in the masses of the $\bar D$ mesons as compared to the
masses of the $D$ mesons, as illustrated in figures \ref{fig9}
and \ref{fig10}.

\begin{figure}
\includegraphics[width=16cm,height=16cm]{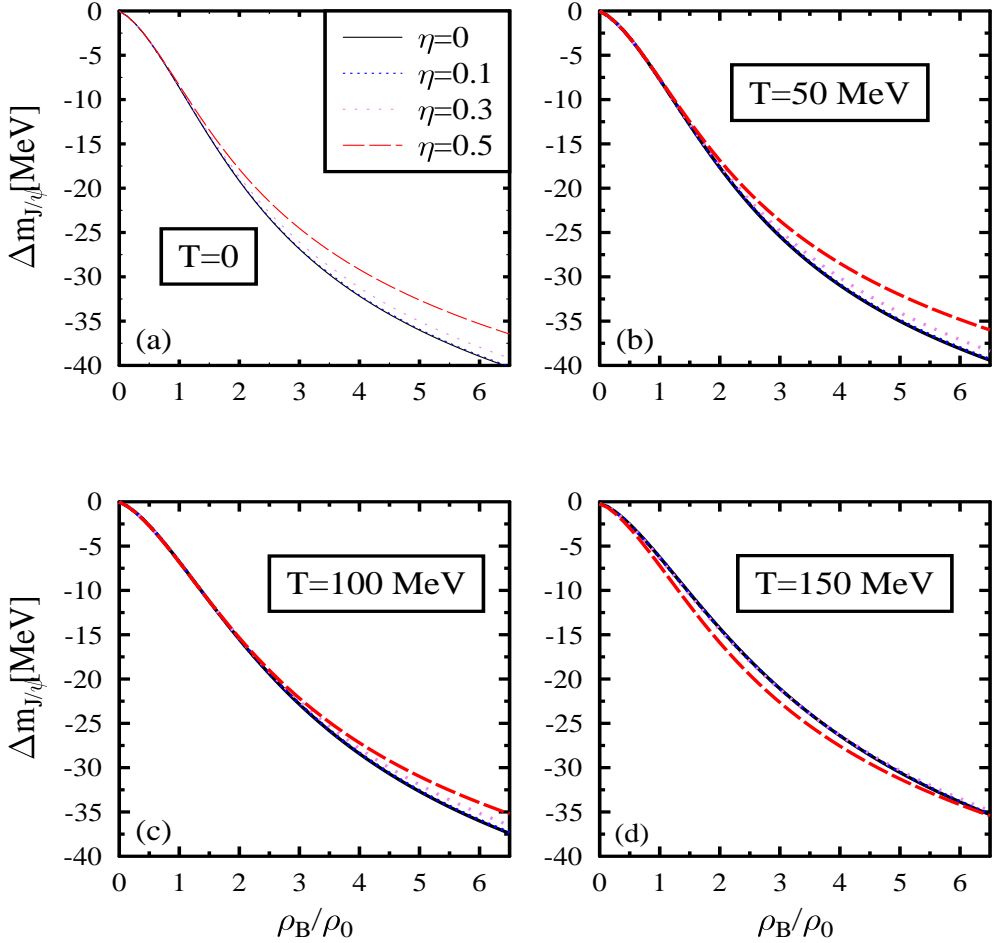} 
\caption{(Color online) The mass shift of $J/\psi$ plotted as a function 
of the baryon density in units of nuclear matter saturation density at 
given temperature, for different
values of the isospin asymmetry parameter, $\eta$.} 
\label{fig15}
\end{figure}

\begin{figure}
\includegraphics[width=16cm,height=16cm]{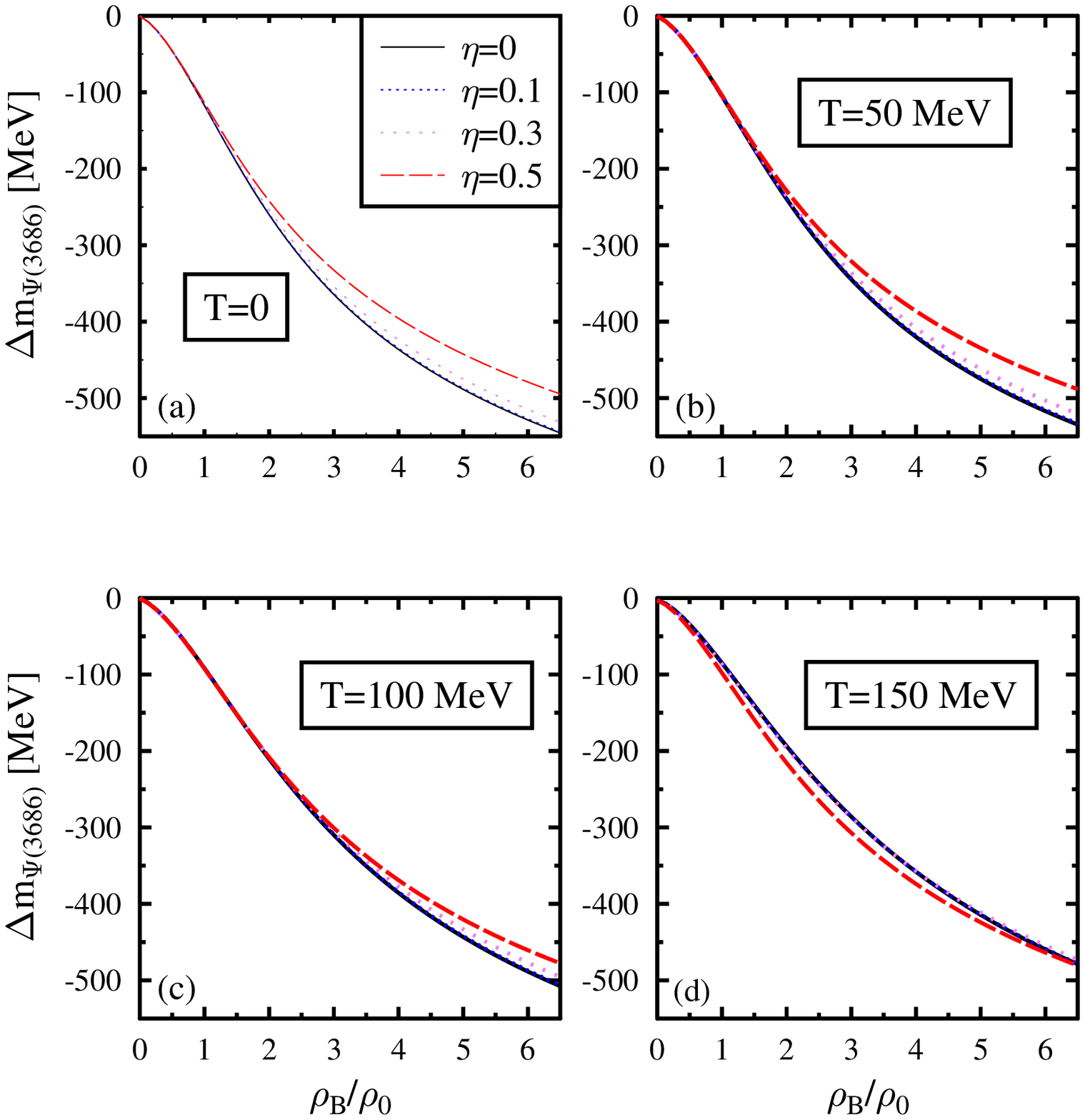} 
\caption{(Color online) The mass shift of $\psi(3686)$ plotted 
as a function of the baryon density in units of nuclear matter
saturation density at given temperature, for different
values of the isospin asymmetry parameter, $\eta$.} 
\label{fig16}
\end{figure}

\begin{figure}
\includegraphics[width=16cm,height=16cm]{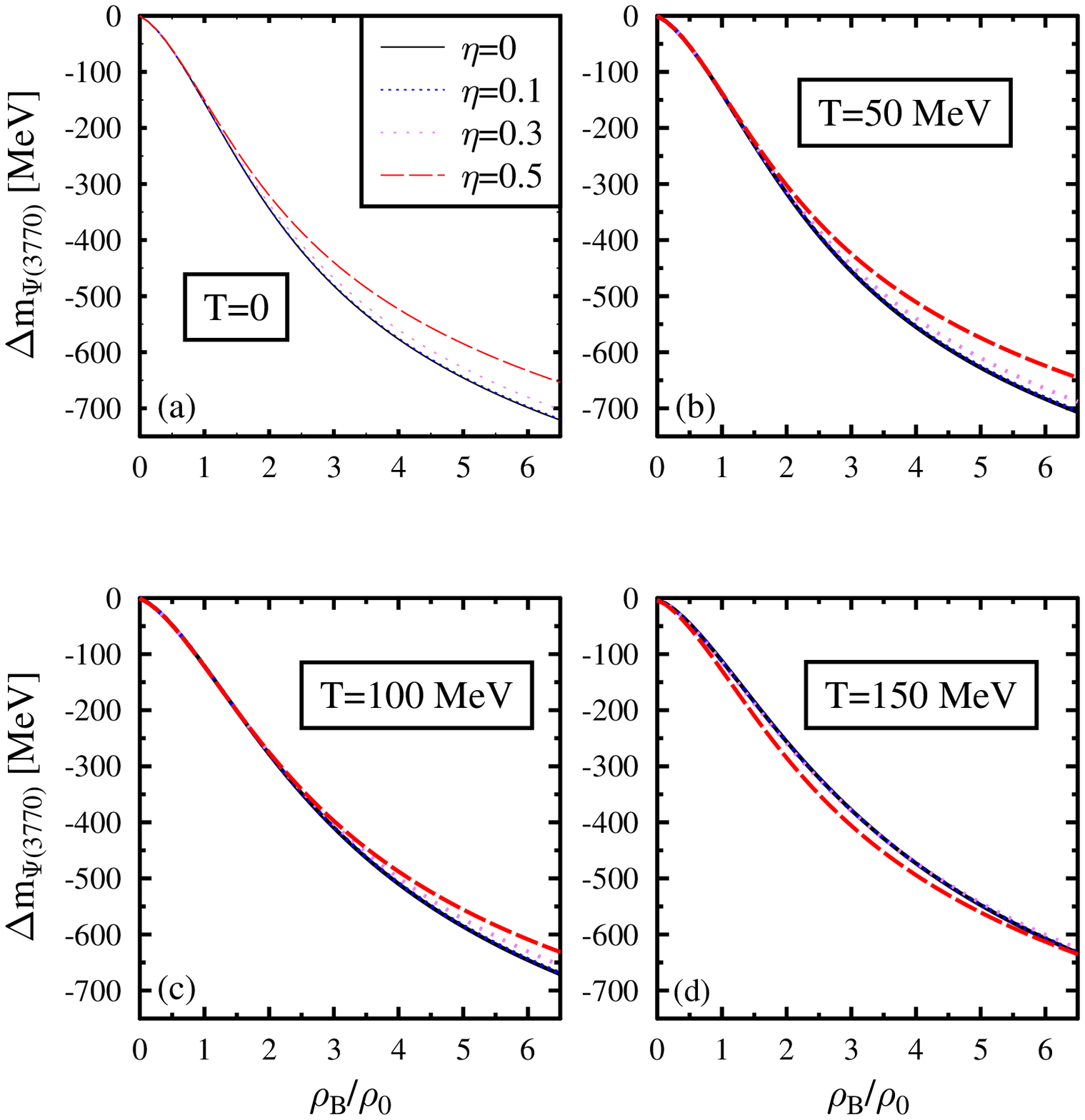} 
\caption{(Color online) The mass shift of $\psi(3770)$ plotted as 
a function of the baryon density in units of nuclear matter saturation
density at given temperature, for different
values of the isospin asymmetry parameter, $\eta$.} 
\label{fig17}
\end{figure}

 We shall now investigate how the behaviour of the dilaton field $\chi$ 
in the hot asymmetric nuclear matter affects the in-medium masses of 
the charmonium states $J/\psi, \psi(3686)$ and $\psi(3770)$. In 
figures \ref{fig15}, \ref{fig16} and \ref{fig17}, we show the shifts 
of the masses of charmonium states $J/\psi, \psi(3686)$ and $\psi(3770)$
from their vacuum values, as functions of the baryon  density for given
values of temperature T and for different values of the isospin asymmetry
parameter, $\eta$. We have shown the results for the values of the 
temperature, T = 0, 50, 100 and 150 MeV. At the nuclear matter saturation 
density, $\rho_{B} = \rho_{0}$ at temperature T = 0, the mass-shift 
for $J/\psi$ meson is $-8.6$ MeV in the isospin symmetric nuclear 
medium ($\eta = 0$) and in the asymmetric nuclear medium, with 
isospin asymmetry parameter $\eta = 0.5$, it is seen to be about $-8.4$ MeV. 
For $\rho_{B} = 4\rho_{0}$ and at zero temperature, the mass-shift for $J/\psi$ 
meson is observed to be about $-32.2$ MeV in the isospin symmetric 
nuclear medium ($\eta = 0$) and in isospin asymmetric nuclear medium 
($\eta = 0.5$), it is seen to be modified to 
$-29.2$ MeV. The increase in the magnitude of the mass-shift, with 
density $\rho_{B}$, is because of the large drop in the dilaton field 
$\chi$ at higher densities. However, with increase in the isospin asymmetry 
of the medium the magnitude of the mass-shift decreases because the drop 
in the dilaton field $\chi$ is less at a higher value of the isospin asymmetry 
parameter $\eta$.
For the nuclear matter saturation density $\rho_{B} = \rho_{0}$ and at 
temperature T = 0, the mass-shift for $\psi(3686)$ is observed to be 
about $-117$ and $-114$ MeV for $\eta = 0$ and $0.5$ respectively, 
and for $\psi(3770)$, the values of the mass-shift are seen to be 
about $-155$ MeV and $-150$ MeV respectively. At $\rho_{B} = 4\rho_{0}$ 
and zero temperature, the values of the mass-shift for $\psi(3686)$ 
are modified to $-436$ MeV and $-396$ MeV for $\eta = 0$ and $0.5$ 
respectively, and, for $\psi(3770)$, the drop in the masses are about
 $-577$ MeV and $-523$ MeV respectively.
As mentioned earlier, the drop in the dilaton field, $\chi$, at finite 
temperature is less than at zero temperature and this behaviour is 
reflected in the smaller mass-shift of the charmonium states at finite 
temperatures as compared to zero temperature case. At nuclear matter 
saturation density $\rho_{B} = \rho_{0}$, temperature T = 100 MeV, 
the values of the mass-shift for the $J/\psi$ meson are observed to be 
about $-6.77$ MeV and $-6.81$ MeV for isospin symmetric ($\eta = 0$) 
and isospin asymmetric ($\eta = 0.5$) nuclear medium respectively. 
At baryon density $\rho_{B} = 4\rho_{0}$, temperature T = 100 MeV, 
the mass-shift for $J/\psi$ is observed to be $-28.4$ MeV and $-27.2$ MeV 
for isospin symmetric ($\eta = 0$) and isospin asymmetric ($\eta = 0.5$) 
nuclear medium respectively. For the excited charmonium states $\psi(3686)$ 
and $\psi(3770)$,  the mass-shifts at nuclear matter saturation density 
$\rho_{B} = \rho_{0}$ and temperature T = 100 MeV, are observed to be 
$-91.8$ MeV and $-121.4$ MeV respectively, for isospin symmetric nuclear 
medium ($\eta = 0$) and $-92.4$ MeV and $-122$ MeV for the isospin 
asymmetric nuclear medium with $\eta = 0.5$. For a baryon density of 
$\rho_{B} = 4\rho_{0}$, and temperature T = 100 MeV, the mass-shifts for 
charmonium states $\psi(3686)$ and $(\psi(3770))$ are seen to be $-386$ MeV 
and $-510$ MeV respectively, for isospin symmetric ($\eta = 0$) and 
$-369$ MeV and $-488$ MeV for isospin asymmetric nuclear medium with 
$\eta = 0.5$. 
For temperature $T = 150$ MeV and at the nuclear matter saturation 
density $\rho_{B} = \rho_{0}$, the mass-shifts for the charmonium 
states $J/\psi, \psi(3686)$ and $\psi(3770)$ are seen to be $-6.25$, 
$-85$ and $-112$ MeV respectively in the isospin symmetric nuclear 
medium ($\eta = 0$). These values are modified to $-7.2$, $-98$ and 
$-129$ MeV respectively, in the isospin asymmetric nuclear medium 
with $\eta = 0.5$. At a baryon density $\rho_{B} = 4\rho_{0}$, 
the values of the mass-shift for $J/\psi, \psi(3686)$ and $\psi(3770)$ 
are observed to be $-26.4$, $-358$ and $-473$ MeV in isospin symmetric 
nuclear medium ($\eta = 0$) and in isospin asymmetric nuclear medium 
with $\eta = 0.5$, these values are modified to $-27.6$, $-375$ and 
$-494$ MeV respectively. Note that at high temperatures, e.g., at 
$T = 150$ MeV the mass-shift in isospin asymmetric nuclear medium 
($\eta = 0.5$) is more as compared to isospin symmetric nuclear medium 
($\eta = 0$). This is opposite to what is observed for the zero temperature 
case. The reason is that at high temperatures the dilaton field $\chi$ has 
a larger drop in the isospin asymmetric nuclear medium ($\eta = 0.5$) as 
compared to the isospin symmetric nuclear medium ($\eta = 0$), due to 
the contributions from the $\delta$ field for the nonzero $\eta$, 
which is observed to decrease in its magnitude at high temperatures.
The dependence of the wave function of the charmonium on the
density and temperature of the medium can be introduced 
through the modification of the strength of the harmonic potential 
for the charmonium state \cite{friman} given as the parameter $\beta$ 
in equation (30). In the medium, one expects the strength
of the confining potential to be smaller than in the vacuum
as due to medium modifications of the hadrons, more decay channels 
can become accessible, which are not available in the vacuum.
We observe that when the parameter is decreased by about 5$\%$,
the mass drops of the charmonium states are increased by
about 14$\%$ and a change of the parameter $\beta$ by 2$\%$
leads to a mass drop of the charmonium states larger
by about 5$\%$.

The values of the mass-shift for the charmonium states obtained within 
the present investigation, at nuclear matter saturation density $\rho_{0}$ 
and temperature $T =0$, are in good agreement with the mass shift of $J/\psi, 
\psi(3686)$ and $\psi(3770)$ as $-8, -100$ and $-140$ MeV respectively, 
at the nuclear matter saturation density computed in Ref. \cite{leeko}
from the second order stark effect, with the gluon condensate in the 
nuclear medium computed in the linear density approximation.
On the other hand, in the present work, the temperature 
and density dependence of the gluon condensates are calculated from 
the medium modifications of the dilaton field within the chiral SU(3) model.  
In Ref. \cite{leeko}, the masses of the charmonium states were calculated 
for the symmetric nuclear matter at zero temperature, whereas the 
present investigation studies the isospin asymmetry dependence of 
the masses of the charmonium states in the nuclear medium at finite 
temperatures, which will be relevant for the asymmetric heavy ion collision 
experiments planned at the future facility at GSI.
The mass-shift for $J/\psi$ has also been studied with the QCD sum 
rules in \cite{klingl} and the value at nuclear saturation density 
was observed to be about $-7$ MeV. 
In \cite{kimlee} the operator product expansion was carried out upto 
dimension six and mass shift for $J/\psi$ was calculated to be $-4$ MeV 
at nuclear matter saturation density $\rho_{0}$ and  at zero temperature. 
The effect of temperature on the $J/\psi$ in deconfinement phase was 
studied in \cite{leetemp, cesa}. In these investigations, it was reported 
that $J/\psi$ mass remains essentially constant within a wide range of 
temperature and above a particular value of the temperature, T, 
there is seen to be a sharp change in the mass of $J/\psi$ 
in the deconfined phase. For example, in Ref. \cite{lee3} 
the mass shift for $J/\psi$ was reported to be about 
200 MeV at T = 1.05 T$_{c}$. In the present work, we have studied the 
effects of temperature, density and isospin asymmetry, on the mass 
modifications of the charmonium states ($J/\psi, \psi(3686)$ and 
$\psi(3770)$) in the confined hadronic phase, arising due to modifications 
of a scalar dilaton field which simulates the gluon condensates of QCD, 
within a chiral SU(3) model. The effect of temperature is found to be 
small for the charmonium states $J/\psi(3097)$, $\psi(3686)$ and 
$\psi(3770)$, whereas the masses of charmonium states are observed to 
vary considerably with density, in the present investigation.

\begin{figure}
\includegraphics[width=16cm,height=16cm]{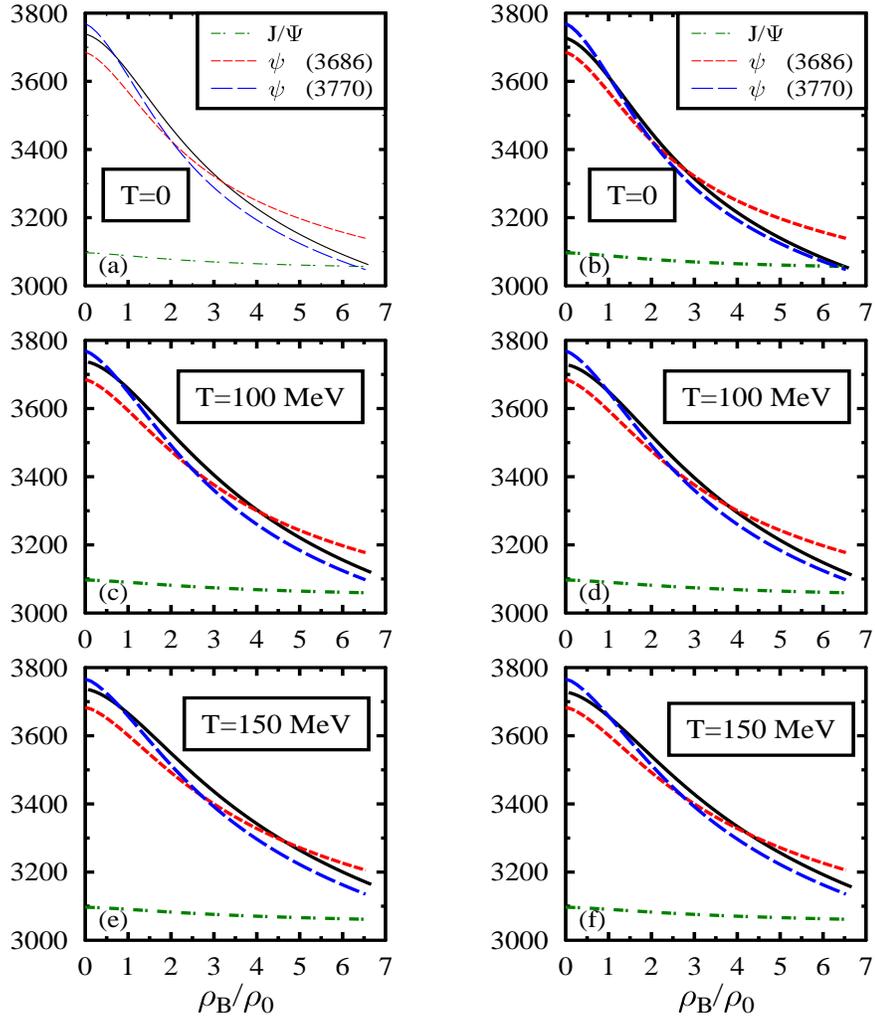} 
\caption{(Color online) The masses of the $D\bar{D}$ pairs 
[$D^{+}D^{-}$ in (a), (c), (e) and $D^{0}\bar{D}^{0}$ in (b), (d), (f)] 
in MeV plotted as functions of $\rho_{B}/\rho_{0}$ for isospin 
for the symmetric nuclear matter ($\eta$=0) and for temperatures, 
T = $0, 100, 150$ MeV.} 
\label{fig18}
\end{figure}

\begin{figure}
\includegraphics[width=16cm,height=16cm]{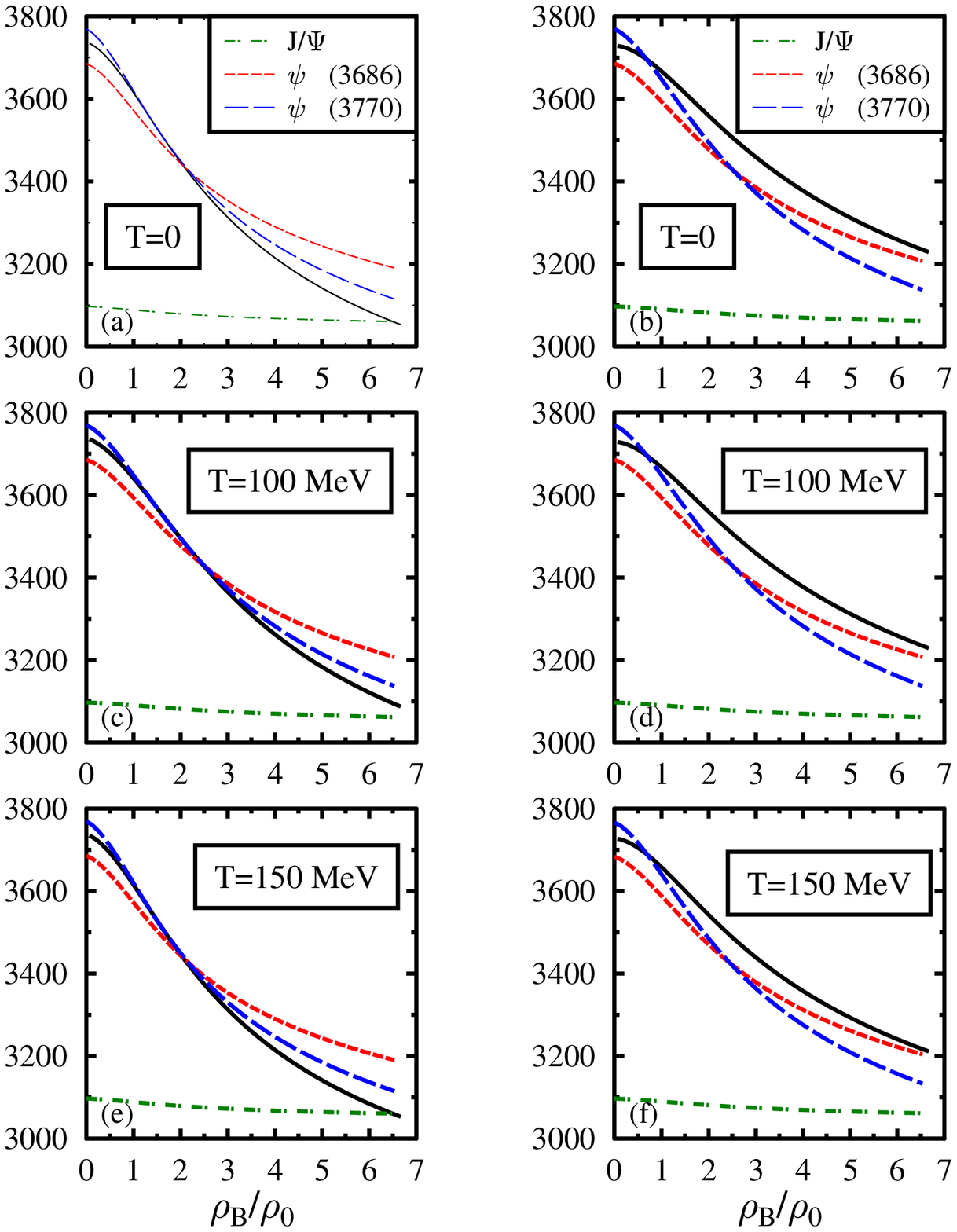} 
\caption{(Color online) The masses of the $D\bar{D}$ pairs 
[$D^{+}D^{-}$ in (a), (c), (e) and $D^{0}\bar{D}^{0}$ in (b), (d), (f)] 
in MeV plotted as functions of $\rho_{B}/\rho_{0}$ for isospin 
asymmetry parameter value $\eta = 0.5 $  and temperatures, 
T = $0, 100, 150$ MeV.}
\label{fig19}
\end{figure}

The medium modifications of the masses of $D$ and $\bar {D}$ mesons 
as well as that of charmonium states could be 
an explanation for the observed $J/\psi$ suppression observed by NA50 
collaboration at $158$ GeV/nucleon in the Pb-Pb collisions \cite{blaiz}. 
Due to the drop in the mass of the $D\bar D$ pair in the nuclear medium,
it can become a possibility that the excited states of charmonium 
($\psi^{'}, \chi_{c2}, \chi_{c1}, \chi_{c0}$) can decay to $D\bar{D}$
pairs \cite{amarind} and hence the production of $J/\Psi$ from the decay of 
these excited states can be suppressed. Even at high values of densities 
at given temperatures, it can become a possibility that $J/\psi$ itself
decays to $D\bar{D}$ pairs. Thus the medium modifications of the $D$ mesons 
can modify the decay widths of the charmonium states \cite{friman}. 
In figures \ref{fig18} and \ref{fig19}, we show the density dependence 
of the masses of the $D^{+}D^{-}$ as well as $D^{0}\bar{D^{0}}$ pairs 
calculated in the present investigation, for temperatures, T = 0, 100 
and 150 MeV and for isospin asymmetry parameter, $\eta = 0$ and
$\eta$=0.5 respectively. We also show the in-medium masses of the charmonium 
states $J/\psi, \psi(3686)$ and $\psi(3770)$ in these figures. 
We observe that in the isospin symmetric nuclear medium at zero 
temperature, the in-medium mass of charmonium $\psi(3770)$ is less than 
$D^{+}D^{-}$ and $D^{0}\bar{D^{0}}$ pairs above baryon densities 
$0.6\rho_{0}$ and $0.8\rho_{0}$ respectively and therefore its 
decay does not seem possible at densities higher than these densities 
in the nuclear medium. 
However, as we move to the isospin asymmetric medium ($\eta = 0.5$), 
the medium modifications for the masses of the $D\bar D$ pairs
as well as of the charmonium states indicate that the decay of 
$\psi(3770)$ to $D^{+}D^{-}$ pairs can be possible above a density 
of about $2\rho_{0}$, but the decay to $D^{0}\bar{D^{0}}$ does  
not seem possible above a density of about nuclear matter saturation
density. In the isospin symmetric nuclear medium, 
the decay of the charmonium state $\psi(3686)$ to $D^{+}D^{-}$
and to $D^{0}\bar{D^{0}}$ seem possible above densities
of about 3.4 $\rho_0$ and 3.3 $\rho_0$ respectively.
In isospin asymmetric nuclear medium ($\eta = 0.5$) 
the decay of the charmonium state $\psi (3686)$ to $D^{+}D^{-}$ 
pairs seem as possibilities above a density of about $2 \rho_{0}$.
The effects of the temperatures on the decay of the 
charmonium states to $D\bar D$ pairs have also been illustrated 
in the figures \ref{fig18} and \ref{fig19}. 
The decay of $\psi (3770)$ to the $D\bar D$ pairs do not seem
possible above a density of about $\rho_0$ even at T=100 and 150 MeV.
However, for $\psi (3686)$, for T=100 MeV, the decay to the $D^+ D^-$
and $D^0 \bar {D^0}$ seem possible above densities of about 4$\rho_0$ 
and 3.8 $\rho_0$ for symmetric nuclear matter
For T=150 MeV, these values are modified to
4.5$\rho_0$ and 4.3 $\rho_0$ respectively, for $\eta$=0.
As we move to the asymmetric nuclear matter, the densities 
above which the decay of $\psi (3686)$ decaying
to $D^+ D^-$ becomes possibile are 2.4 $\rho_0$ and 2.2 $\rho_0$
respectively.  Similar to the zero temperature case, we do not
see a possibility of $\psi (3686)$ to $D^0 \bar {D^0}$ 
for the asymmetric nuclear matter ($\eta$=0.5) at T=100
and 150 MeV. In the present investigation, the decay of $J/\psi$ 
to $D\bar D$ pairs does not seem as a possibilty even upto a density 
of about 6$\rho_0$. We observe from figures 18 and 19 that the 
temperature dependence is minimal for the decay of the charmonium
states to the $D\bar D$ pairs even though the density dependence
is quite appreciable. 

The decay of the charmonium states have been studied in Ref. 
\cite{friman,brat6}. It is seen to depend sensitively on the relative 
momentum in the final state. These excited states might become narrow 
\cite{friman} though the $D$ meson mass is decreased appreciably at 
high densities. It may even vanish at certain momentum corresponding 
to nodes in the wave function \cite{friman}. Though the decay widths 
for these excited states can be modified by their wave functions, the 
partial decay width of $\chi_{c2}$, owing to absence of any nodes, 
can increase monotonically with the drop of the $D^{+}D^{-}$ pair mass 
in the medium. This can give rise to depletion in the  $J/\Psi$ yield 
in heavy-ion collisions. The dissociation of the quarkonium states 
($\Psi^{'}$,$\chi_{c}$, $J/\Psi$) into $D\bar{D}$ pairs has also been 
studied \cite{wong,digal} by comparing their binding energies with the 
lattice results on the temperature dependence of the heavy-quark 
effective potential \cite{lattice}.

\section{summary}
We have investigated in a chiral model the in-medium masses of the $D$, 
$\bar{D}$ mesons and the charmonium states ($J/\psi$, $\psi (3686)$
and $\psi (3770)$) in hot isospin asymmetric nuclear matter. The properties 
of the light hadrons -- as studied in  $SU(3)$ chiral model -- modify the 
$D(\bar{D})$ meson properties in the dense and hot hadronic matter. The 
$SU(3)$ model, with parameters fixed from the properties of the hadron 
masses in vacuum and low-energy KN scattering data, is extended to 
SU(4) to derive the interactions of $D(\bar{D})$ mesons with the light 
hadron sector. The mass modifications of $D^{+}$ and $D^{0}$ mesons is 
strongly dependent on isospin-asymmetry of medium when we determine
the parameters $d_1$ and $d_2$ consistent with the KN scattering lengths.
However, the sensitivity to the isospin asymmetry is seen to be more
for the $\bar D$ doublet, when we fit the parameters to the DN scattering 
lengths as calculated in the coupled channel approach in Ref. \cite{MK}. 
At finite densities, the masses of $D (\bar D)$ mesons are observed 
to increase with temperature \cite{amdmeson} upto a temperature 
above which it is observed to decrease. The mass modification for the $D$ 
mesons are seen to be similar to earlier finite density calculations of 
QCD sum rules \cite{qcdsum08,weise} as well as to the quark-meson coupling 
model \cite{qmc}. This is in contrast to the small mass modifications 
in the coupled channel approach \cite{ljhs,mizutani8}. Also we obtain 
small attractive mass shifts for $\bar{D}$ mesons similar to the results 
obtained from the QMC model, which might lead to formation of charmed 
mesic nuclei. These results for the $\bar D$ mesons are contrary 
to the results from the coupled channel approach \cite{mizutani8}, 
where the $\bar D$ mesons experience a repulsive interaction in the 
nuclear medium. In our calculations the presence of the 
repulsive  first range term (with coefficient $-\frac{1}{f_{D}}$ in 
Eq. (\ref{ldn})) is compensated by the attractive $d_{1}$ and $d_{2}$ 
terms  in Eq.(\ref{ldn}). Among the attractive range terms ($d_{1}$ and 
$d_{2}$ terms), $d_{1}$ term is found to be dominating over $d_{2}$ term.

We have investigated in the present work, the effects of density, 
temperature and isospin asymmetry of the nuclear medium on the masses 
of the charmonium states $J/\psi$, $\psi(3686)$ and $(\psi(3770)$, 
arising due to modification of the scalar dilaton field, $\chi$, 
which simulates the gluon condensates of QCD, within the chiral 
SU(3) model. The change in the mass of $J/\psi$ with the density 
is observed to be small at nuclear matter saturation density and 
is in agreement with the QCD sum rule calculations. There is seen 
to be appreciable drop in the in-medium masses of excited charmonium 
states $\psi(3686)$ and $\psi(3770)$ with density. The mass drop 
of the excited charmonium states $\psi(3686)$ and  $\psi(3770)$ 
are large enough to be seen in the dilepton spectra emitted from 
their decays in experiments involving $\bar{p}$-A annihilation 
in the future facility at GSI, provided these states decay 
inside the nucleus. The life time 
of the $J/\psi$ has been shown to be almost constant in the nuclear 
medium, whereas for these excited charmonium states, the
life times are shown to reduce to less than 5 fm/c, due to appreciable 
increase in their decay widths \cite{friman}. Hence a significant 
fraction of the produced excited charmonium states in these experiments 
are expected to decay inside
the nucleus \cite{golu}. The in-medium properties of the excited 
charmonium states $\psi(3686)$ and $\psi(3770)$ can be studied 
in the dilepton spectra in $\bar{p}$-A experiments in the future 
facility of the FAIR, GSI \cite{gsi}. The mass shift of the charmonium 
states in the hot nuclear medium seem to be appreciable at high densities 
as compared to the temperature effects on these masses, and these should 
show in observables like the production of these charmonium states in the
compressed baryonic matter experiment at the future facility at GSI, 
where baryonic matter at high densities and moderate temperatures will
be produced.
 
The medium modifications of the $D$ meson masses can lead to a suppression 
in the $J/\Psi$ yield in heavy-ion collisions, since the excited states of 
the $J/\Psi$ can decay to $D\bar{D}$ pairs in the dense hadronic 
medium. The medium modifications of the masses of
the charmonium states as well as the $D$ and $\bar{D}$ mesons have been
considered in the present investigation. The isospin asymmetry lowers 
the density at which decay to $D^{+}D^{-}$ pairs occur. Due to increase 
in the mass of $D^{0}\bar{D}^{0}$ in the isospin-asymmetric medium, 
isospin-asymmetry is seen to disfavor the decay of the charmonium states 
to the $D^{0}\bar{D}^{0}$  pairs. At zero or
finite temperatures, there does not seem to be a possibility of decay 
of $J/\Psi$ to $D^+ D^-$ or  $D^{0}\bar{D}^{0}$  pairs. 
The isospin dependence of 
$D^{+}$  and $D^{0}$ masses is seen to be a dominant medium effect 
at high densities, which might show in their production ($D^{+}/D^{0}$), 
whereas, for the $D^{-}$ and $\bar{D}^{0}$, one sees that, even though 
these have a strong density dependence, their in-medium masses remain 
similar at a given value for the isospin-asymmetry parameter $\eta$. 
This is the case when we fit the parameters $d_1$ and $d_2$ from
the KN scattering lengths. When we determine these parameters
from the DN scattering lengths as calculated in Ref. \cite{MK},
the masses of the $\bar D$ doublet are seen to be more sensitive
to isospin asymmetry in the medium.
The strong density dependence as well as the isospin dependence of 
the $D(\bar{D})$ meson optical potentials in asymmetric nuclear matter 
can be tested in the asymmetric heavy-ion collision experiments at 
future GSI facility \cite{gsi} in observables like the
$D^+/D^0$ as well as $D^-/\bar {D^0}$ ratios. In the present work,
we have investigated the in-medium masses of the charmonium states
due to their interaction with the scalar dilaton field (simulating
the gluon condensates of QCD) and $D (\bar D)$ mesons due to the
interaction with nucleons as well as scalar mesons.
The parameters of the asymmetric nuclear matter are fitted to the
nuclear matter properties, vacuum baryon masses, the hyperon potentials 
as well as the KN scattering lengths in the chiral SU(3) model.
The study of the in-medium modifications of $D$ mesons in hadronic matter
including hyperons along with nucleons at zero and finite temperatures,
as well as the study of the medium modifications of the strange 
open charm mesons can be possible extensions of the present 
investigation.

\acknowledgements
Financial support from Department of Science and Technology, Government 
of India (project no. SR/S2/HEP-21/2006) is gratefully acknowledged 
by the authors. One of the authors (AM) is grateful to the Frankfurt
Institute of Advanced Studies (FIAS), University of Frankfurt, 
for warm hospitality and acknowledges financial support from 
Alexander von Humboldt Stiftung when this work was initiated.

\end{document}